\documentclass[12pt]{article}
\usepackage{titletoc}
\usepackage[toctitles]{titlesec}
\usepackage[toc,page]{appendix}
\usepackage{tocloft}
\usepackage{amssymb,amsmath,amsthm}
\usepackage{natbib}
\usepackage[margin=1in]{geometry}
\usepackage{graphicx}	
\usepackage{setspace}
\usepackage{dcolumn}
\usepackage{comment} 
\usepackage{lscape}
\usepackage{footmisc} 
\usepackage{multirow}
\usepackage{systeme}
\usepackage{tikz} 
\usetikzlibrary{patterns} 
\usepackage{sgame} 
\usepackage{color}
\usepackage{float}
\usepackage[colorlinks=true,allcolors=blue]{hyperref}  
\usepackage{mathtools}
\usepackage[final]{pdfpages}

\theoremstyle{plain}

\newtheorem{theorem}{Theorem}

\newtheorem{lemma}{Lemma}[section]

\theoremstyle{definition}
\newtheorem{example}{Example}

\newcommand{\crs}{corresponding }

\newcommand{\eqm}{equilibrium }
\newcommand{\eqst}{equilibrium system }

\newcommand{\crsp}{corresponding }

\newcommand{\dst}{distribution }

\newcommand{\flw}{following}

\newcommand{\mpt}{important }
\newcommand{\lbls}{liabilities }

\newcommand{\nvt}{invariant }

\newcommand{\blgs}{obligations}
\newcommand{\prt}{partition }

\newcommand{\prbs}{probabilities }

\newcommand{\sth}{stochastic }

\newcommand{\trt}{transient }

\makeatother

\begin{document}

\author{Isaac M. Sonin \\
%EndAName
University of NC at Charlotte \and Konstantin Sonin \\
%EndAName
University of Chicago}
\title{Banks as Tanks: A Continuous-Time Model of Financial Clearing\textbf{\thanks{%
We thank Daron Acemoglu, Michael Grabchak, Yury Kabanov, and Ernst Presman for their very helpful comments.}}}

\date{\today }
\maketitle

\thispagestyle{empty}\bigskip \bigskip

\begin{small}%

\vspace{-0.3in} %% real mechanism of simultaneous payments

\begin{center}
\textbf{Abstract}
\end{center}

\baselineskip0.5cm

{\small We present a simple continuous-time model of clearing in financial networks. Financial firms are represented as ``tanks'' filled with fluid (money), flowing in and out. Once ``pipes'' connecting ``tanks'' are open, the system reaches the clearing payment vector in finite time. This approach provides a simple recursive solution to a classical static model of financial clearing in bankruptcy, and suggests a practical payment mechanism. With sufficient resources, a system of mutual obligations can be restructured into an equivalent system that has a cascade structure: there is a group of banks that paid off their debts, another group that owes money only to banks in the first group, and so on. Technically, we use the machinery of Markov chains to analyze evolution of a deterministic dynamical system.}

\bigskip




\bigskip

%TCIMACRO{\TeXButton{EndSmall}{\end{small}}}%
%BeginExpansion
\end{small}%
%EndExpansion

\bigskip \newpage

\newpage

\tableofcontents
\thispagestyle{empty}

\clearpage

\setcounter{page}{1}

\baselineskip0.71cm

\section{Introduction\label{Int}}

Modern financial systems are richly interconnected networks of various
institutions, which depend on each other. With most of assets of one firm
being the liabilities of other financial institutions, such a system is
naturally prone to \textquotedblleft systemic risk\textquotedblright\ of
contagions, in which a default of one firm might lead to defaults of many others (\citealp{allen2000,glasserman2016}). Optimal regulations of such networks is a subject of
intensive debate \citep[e.g.,][]{yellen2013interconnectedness}.

In a seminal contribution, \cite{eisenberg2001systemic} introduced a framework
for study of interlinked financial systems in distress. They considered an environment, in
which all participants (henceforth, \textquotedblleft
banks\textquotedblright) default within a single clearing mechanism, and
demonstrated that there always exist a
\textquotedblleft clearing payment vector\textquotedblright\ that satisfy
some natural requirements. The Eisenberg-Noe
approach has been successfully extended to incorporate liquidity spillovers (\citealp{cifuentes2005,SHIN2008}), outside liabilities (\citealp{Elsinger2009,GLASSERMAN2015}), costs of default (\citealp{rogers2013}), liabilities of different seniority (\citealp{kusnetsov2019interbank}), mandatory disclosures (\citealp{alvarez2015}), and other financial instruments, and has become a cornerstone in analysis of systemic financial risk \citep[see, e.g.,][for recent surveys]{hurd2016contagion,Feinstein2018,kabanov2018clearing}.

In this paper, we develop a continuous-time model of financial clearing. We
think of interconnected firms as \emph{tanks} filled with money, which flow
in and out through pipes connecting the tanks. In the case of default, the
pipes open and, given the assumptions about the intensity of the flows
similar to assumptions that of \cite{eisenberg2001systemic}, the resulting
distribution of liquidity in tanks correspond to the clearing payment vector. We fully characterize the evolution of the dynamical system, which in turns allows to calculate the payment schedule.

Second, the evolution of continuous-time model provides an intuitive explanation for the possible multiplicity of clearing payments vectors. At each moment of time, we interpret the banking network as a finite Markov chain -- despite that our model is deterministic -- and study its evolution using the stochastic matrix of current relative liabilities. Groups of banks that may be involved in mutual reduction of debts without any real money transfers, which we call ``swamps'', correspond to ergodic subclasses of the invariant distribution of the Markov chain.\footnote{Though swamps might look like a mathematical abstraction, they do appear in the real world. For example, the infamous \emph{Enron} company created special purpose entities that conducted circular transactions, inflating the company's revenues and hiding the costs \citep{Healy2003}.} It is these ``swamps'' that are responsible for non-uniquenesses of payment schedules leading ultimately to the same clearing payment vector.

Third, our approach provides a simple, but intuitive method to reduce many networks of mutual liabilities to an equivalent cascade network, in which there is a group of banks that do not have any obligations, the second group of banks that owe money to banks of the first group only, and so on. In these circumstances, there always exists a bank that will repay its obligations in full, even if \emph{all} banks start with a negative amount of cash on hand (outside liabilities). Restructuring that reduces the amount of obligations is practically important, as growing complexity of financial networks creates a significant hurdle for regulators (\citealp{yellen2013interconnectedness,bernard2017}).

The model demonstrates that the clearing mechanism does not require sophisticated planning and hands-on management on behalf of the regulators. In fact, allowing financially-constrained firms to repay their debts at the maximum-available speed without any liquidity injections will result in the clearing
payment vector. While the existing clearing mechanisms address a number of
important issues, our results show that, at least theoretically, there is no
need in detailed regulation in a situation of financial distress. This is especially important given the growing recognition of challenges that regulators face dealing with extensive and complicated networks of mutual liabilities.\footnote{\cite{glasserman2016} cite FCIC report: “There was no way to know who would be owed how much and when payments would have to be made—information that would be critically important to analyze the possible impact of a Lehman bankruptcy on derivatives counterparties and the financial markets.”}

Finally, our results demonstrate the power of continuous-time approach to
static maximization problems. Our continuous-time model generates a discrete dynamical system with the following recursive structure: at each moment, there is a linear system of equation that determines the parameters of the next state of the system and the moment when the system switches to this new state. In a finite number of steps, the system reaches its ultimate state. Thus, the continuous-time model allows to easily calculate the payment schedule.

An important advantage of the dynamical model is that it allows to study different aspects of possible interaction between banks. The model can be extended to deal with financial obligations of multiple maturities in real time, and to be used to take into account stochasticity. It might be a convenient tool to study the optimal strategy of a central agency to minimize the potential contagion effect triggered by failure of some banks. In particular, it is straightforward to calculate the minimum amount of cash to be injected into interconnected system to have all banks fully paid off.

\cite{allen2000} and \cite{freixas2000} pioneered studies of the role of interconnectedness in stability of financial networks. \cite{rogers2013} extended \cite{eisenberg2001systemic} by introducing the possibility of a partial default, and described an algorithm that results in the largest clearing vector for any environment (see also \citealp{AminiFilipovic2016}). \cite{BATTISTON2012} demonstrated that a higher diversification of each individual network member might lead to an increase, rather than a decrease, of a systemic risk. \cite{LIU2010}  provided a formula for sensitivity analysis of Eisenberg and Noe’s one-period model of contagion via direct bilateral links. \cite{Feinstein2018} quantified the Eisenberg--Noe clearing vector's sensitivity.

In an empirical study, \cite{azizpour2018} demonstrated that contagions, through which the defaulting firms have a direct impact on the financial health of their counterparts, is a significant source of default clustering. \cite{Banerjee2000} generalized the model to allow for continuous time dynamics of the interbank liabilities using stochastic differential equations, prove existence of a solution for this generalized model, and analyzed properties of this solution. \cite{csoka2018} used the Eisenberg-Noe approach to analyze decentralized clearing. \cite{capponi2020} applied it to study of optimal policy in the case of a sovereign default.

There is a substantial literature that focuses on specific forms of networks connecting financial firms. \cite{acemoglu2015systemic} demonstrated the non-monotonic effect of the magnitude of negative shocks on network stability. The network density contributes to stability when the shocks are relatively small, but propagate contagion at a certain magnitude. \cite{Castiglionesi2018} show that the star-shaped network is most resilient to systemic risks. In this paper, we take the network as given, and our results apply to all networks. Still, what our model highlights is that a restructuring of a system of mutual obligations might lead to another, simpler system, which is equivalent in terms of the ultimate clearing vector.

\cite{veraart2020} use a network model to describe how contagions, including \emph{distress contagions,} which precede, rather than follow, an institution's bankruptcy, spread through the system. \cite{elliot2014} analyze the role of integration and diversification in prevention of cascade failures. \cite{amini2016} study the resilience of a large financial network to defaults using contagion in random graphs. \cite{FEINSTEIN2017} generalized \cite{AminiFilipovic2016} by allowing for differing liquidation strategies, and provided sufficient conditions for the existence of an equilibrium liquidation strategy with corresponding unique clearing vector. \cite{cifuentes2005} note that the full connectedness of the financial network is a sufficient condition for the uniqueness of the clearing vector (see also \citealp{shin2010}). Our approach allows to refine this observation: it is the presence of ergodic subclasses among banks that has zero cash at the initial moment that generates multiplicity.

\medskip

The rest of the paper is organized as follows. In Section \ref{ENm}, we
outline the original \cite{eisenberg2001systemic} model and use it to discuss the main questions that our continuous-time model addresses. Section \ref{Setup} introduces the continuous-time setup. In Section \ref{SubsectionRegularCase}, we prove the existence theorem and characterize the dynamic process that leads to the clearing vector.  In Section \ref{Swm}, we use Markov chains to characterize the sources of  multiplicity of clearing vectors. In Section \ref{SectionCascade}, we focus on models that can be restructured into the cascade form.  Section \ref{SectionExtensions} discusses extensions. Section \ref{CON} concludes.

\section{Eisenberg and Noe Framework\label{ENm}}

In this Section, we outline the original \cite{eisenberg2001systemic} model
and use it to discuss the main questions that our continuous-time model
seeks to answer and describe informally the answers our model provides.

\paragraph{Eisenberg and Noe (2001).}

There is a financial system composed of $n$ banks (firms, financial institutions, agents, etc) owing to each other, which is described by the matrix of mutual liabilities $B=\{b_{ij}\}$.
Vector $\mathbf{c}=(c_{1},c_{2},...,c_{n})$ describes the initial cash
positions, $c_i\ge 0$ for all $i$.

In our continuous-time model, we will use an equivalent system of
parameters; in our model, they will be functions of time. Matrix of mutual liabilities $Q$ uniquely determines the \emph{debt vector} $\mathbf{b}=(b_{1},b_{2},...,b_{n})$ and the stochastic \emph{matrix of relative liabilities} $Q=\{q_{ij}\}$, so that $b_{ij}=b_{i}q_{ij}$ for every $i,j$. \emph{Vice versa}, matrix $Q$ and vector $\mathbf{b}$ uniquely determine the matrix of liabilities $B$.

The \textit{clearing vector} $\mathbf{p}=(p_{1},p_{2},...,p_{n})$, where $%
p_{i}$ is the total amount paid by bank $i$ to all other banks, should
satisfy the following two conditions \citep{eisenberg2001systemic}:\vspace{0.2cm}

\noindent \textbf{(A)}\textit{\ Priority and proportionality of debt claims. }With the
total payment, $p_{i}$, the bank $i$ pays to $j$ the fraction $p_{i}q_{ij}$
in such a way that either its total debts are paid or all of its resources
are exhausted.\vspace{0.2cm}

\noindent \textbf{(B)}\textit{\ Limited liability.} The total payment of any bank
should never exceed the cash flow available to the bank, i.e., the initial
cash plus money received from other banks.\vspace{0.2cm}

Conditions (A) and (B), taken together, imply that the clearing vector $%
\mathbf{p}$ satisfies
\begin{equation}
\mathbf{p}=\min (\mathbf{c}+Q^{T}\mathbf{p},\mathbf{b}).  \label{Basic}
\end{equation}
In all equations, $\mathbf{p},\mathbf{c}$, and $\mathbf{b}$ are column $n$%
-dimensional vectors, the minimum is taken component-wise, $T$ is the
transposition symbol, and thus $(Q^{T}\mathbf{p})_{i}$ equals the total
amount received by bank $i$ from other banks. The existence of such a vector
$\mathbf{p}$ follows, e.g., from the Knaster--Tarski lattice version of the
fixed-point theorem, but in our model its existence will follow almost
immediately from the structure of our model. If $p_{i}<b_{i}$ we say that
bank $i$ is in default.

Our Theorem \ref{Main} shows that if the system starts with the debt and cash vectors as in \cite{eisenberg2001systemic}, then, with ``pipes'' open and money flowing between ``tanks'', the outcome is exactly the clearing vector. It also gives a full characterization of a discrete dynamical system that can be solved via a straightforward algorithm.

If all banks start the procedure with some cash on hands, the solution path is unique. Otherwise, there might be complications. Note that condition (A) is a strong requirement: it implies that if, e.g., $b_{ij}=1$, and $b_{ji}=2$, then these banks cannot \textquotedblleft cancel\textquotedblright\ these liabilities by replacing them by $b_{ij}=0$, and $b_{ji}=1$, since the proportionality of debt claims will be violated. Another observation is that even if all $c_{i}=0$, there may exist nontrivial solutions of equation (\ref{Basic}), and we will show that this is exactly the cause of potential multiplicity of solutions.

\paragraph{The Transactions Problem.}

The main focus of the original \cite{eisenberg2001systemic} was on the existence
of the clearing vector $\mathbf{p}$, a solution to the fundamental clearing
equation (\ref{Basic}). Yet even when the clearing vector $\mathbf{p}$ is
known, how the banks will pay their liabilities? How could they do this simultaneously? What is the optimal sequence of payments between banks that result in the clearing vector $%
\mathbf{p}$?

To address these questions, let us introduce a few definitions.
Given a model $M=(\mathbf{b},Q,\mathbf{c})$, we will have, in addition to
the clearing vector satisfies $\mathbf{p}^{\ast }$, i.e., \emph{final payments,} the
vector of \emph{amounts received} $\mathbf{r}^{\ast }=Q^{T}\mathbf{p}^{\ast
} $, the \emph{final cash positions} $\mathbf{c}^{\ast }=\mathbf{c}+\mathbf{r%
}^{\ast }-\mathbf{p}^{\ast }$, the \emph{unpaid debts} $\mathbf{b}^{\ast }=%
\mathbf{b}-\mathbf{p}^{\ast },$ and \emph{losses} $\mathbf{l}^{\ast }=Q^{T}(%
\mathbf{b}-\mathbf{p}^{\ast })$.

We also have the following balance conditions:
\begin{eqnarray}
\sum_{i}c_{i} &=&\sum_{i}c_{i}^{\ast },\   \notag \\
\sum_{i}b_{i} &=&\sum_{i}(Q^{T}b)_{i},\ \   \notag \\
\ \sum_{i}p_{i}^{\ast } &=&\sum_{i}(Q^{T}p^{\ast })_{i},  \label{ge} \\
\sum_{i}l_{i}^{\ast } &=&\sum_{i}(b_{i}-p_{i}^{\ast }).  \notag
\end{eqnarray}%
The first three equations follow since the model is self-contained. The last
equality follows form the previous equations.

Given a \emph{pair} $(\mathbf{b},Q)$, it is useful to introduce the \emph{%
vector of minimum cash positions} $\mathbf{c}^{m}(\mathbf{b},Q)$, the minimum
vector such that all banks should have to be able to pay their debts. In
this case $\mathbf{p}^{\ast }=\mathbf{b}$, $\mathbf{c}^{\ast }=(0,...,0)$
and two terms in the right side of clearing equation (\ref{Basic}) must
coincide, i.e., $\mathbf{b}=\mathbf{c}+Q^{T}\mathbf{p}$.

Let us say that vector $\mathbf{u}=(u_{1},...,u_{n})$ \emph{is larger
(component-wise)} than vector $\mathbf{v}=(v_{1},...,v_{n})$ if $u_{i}\geq
v_{i},$ $i=1,...,n$ and denote $\mathbf{u}\succ \mathbf{v}$.

Using the clearing equation, we immediately obtain simple but useful results about the vector of minimum cash positions.

First, this vector satisfies
\begin{eqnarray}
\mathbf{c}^{m} &=&(I-Q^{T})\mathbf{b},  \label{cm}
\end{eqnarray}

Second,
\begin{equation}
\sum_{i=1}^{n}c_{i}^{m}=0.  \label{Null}
\end{equation}

Third, if the initial cash vector $\mathbf{c}$ satisfies $\mathbf{c}\succ
\mathbf{c}^{m},$ then $\mathbf{c}^{\ast }=\mathbf{c}-\mathbf{c}^{m}$ and the
clearing vector satisfies $\mathbf{p}^{\ast }=\mathbf{b}$. Otherwise, at least one bank will be in default.

In the case when the clearing vector $\mathbf{p}^{\ast }=\mathbf{b}$, we say
that the model is \emph{sufficient}, otherwise \emph{deficient}. Obviously,
vector $\mathbf{c}^{m}$ is the minimum vector that makes the model
sufficient. Thus, with the same parameters $Q$ and $\mathbf{b}$ model will
be sufficient or deficient depending on cash vector $\mathbf{c}$. The
difference between the two cases is substantial: in the former case, the
pairwise mutual obligations can be cancelled, in the latter they cannot. To
answer whether or not vector $\mathbf{c}$ is a sufficient one or a deficient
there is no need to solve clearing equation, we need only to compare this
vector with vector $\mathbf{c}^{m}$ which can be calculated using (\ref{cm}).

\begin{example}
\label{ExampleOne} Let $n=3$, and matrix $Q$\
be given by $q_{12}=q_{13}=\frac{1}{2},q_{21}=\frac{1}{3},q_{23}=\frac{2}{3},%
q_{31}=\frac{1}{4},$ and $q_{32}=\frac{3}{4}:$
\begin{eqnarray}
\label{MatrixQ}
Q=
	\begin{pmatrix}
         $0$ & \quad $1/2$ & \quad$1/2$ \\
	        $1/3$ & \quad $0 $ & \quad$2/3$ \\

         $1/4$ & \quad $3/4$ & \quad$0$ \\
	\end{pmatrix}
\end{eqnarray}
With this matrix, consider the debt vector $\mathbf{b}=(12,21,20).$ For $\mathbf{b}$, using formula (\ref{cm}), we obtain $\mathbf{c}^{m}\left( Q,%
\mathbf{b}\right) =(0,0,0)$. That is, to pay all the debts it is sufficient that no bank has negative cash at the initial moment.
\end{example}

If the cash vector is sufficient, i.e. $\mathbf{c}\succ \mathbf{c}^{m}$,
then the clearing (payment) vectors $\mathbf{p}%
^{\ast }$ coincides with debt vectors $\mathbf{b.}$ In
the case of $\mathbf{b}=(12,21,20)$ in Example \ref{ExampleOne} \emph{no transactions is necessary}. The job
of a clearing house in this case is just to inform the banks that their
obligations cancel each other. A natural question is as follows: given
matrix $Q,$ for which debt vector $\mathbf{b},$ the minimum cash vector is $%
\mathbf{c}^{m}=(0,0,0)$? As we shall see, the necessary and sufficient
condition is that the debt vector $\mathbf{b}$ should be proportional to the
invariant distribution $\pi $ of the corresponding ergodic Markov chain with
matrix $Q$. In Example \ref{ExampleOne}, $\pi =\frac{1}{53}(12,21,20)$. The general result is given in Theorem \ref{Th2}.

\paragraph{Cascade Structure of Liabilities.}

The minimum initial cash vector $\mathbf{c}^{m}$ that satisfies (\ref{cm}) answers the following question. What is the smallest initial cash vector $\mathbf{c}$ such that \emph{all banks} ultimately pay their debts? Now, given matrix $Q$ and vector $b$, what is the smallest vector $\mathbf{c}$ such that \emph{at least one bank} will pay its debt? The answer to this question might seem surprising: a bank that fully repays its debt exists for \emph{any} initial cash vector vector $\mathbf{c}=(c_1,...,c_n)$, even if all $c_i<0$. Indeed, (\ref{Null}) implies that if $c_{i}^{m}>0$ for at least one bank, then for at least one other bank $c_{j}^{m}<0$, i.e., this bank $j$ will ultimately be a net recipient and can start with a negative cash position.
If $\mathbf{c}^{m}=(0,...,0),$ then mutual obligations of all banks cancel
each other, thus no real transactions are necessary.

Let us go through the actual payment mechanism in the next example.
\begin{example}
\label{ExampleTwo}
Consider the same matrix $Q$ given by (\ref{MatrixQ}), but a different debt vector, $\mathbf{b}=(13,22,20).$ Using formula (\ref{cm}), we obtain the minimum cash vector $\mathbf{c}^{m}\left( Q,\mathbf{b}\right) =\left(\frac{2}{3},\frac{1}{2},-\frac{7}{6}\right):$ as we will see, even with a negative initial cash position, bank 3 will be able to pay its debts in full. Indeed, let us demonstrate that debt vector $\mathbf{b}=(13,22,20)$ is equivalent to debt vector $(1,1,0)$. Suppose bank 1 prepares payments of 6 to bank 2 and 6 for banks 3, bank 2 prepares payments of 7 to bank 1 and 14 to bank 3, and bank 3 prepares payments of 5
to bank 1 and 15 to bank 2. These potential payments satisfy the proportionality assumption (A), but formally violate assumption (B). Yet these potential payments partially cancel each other: e.g., 6 from bank 1 to bank 2, and 7 from bank 2 to bank 1 result in just 1 owed by bank 2 to bank 1. Thus, the remaining debt vector is $(1,1,0)$. Given matrix $Q,$ bank 3 has no debt, bank 1 owes $\frac{1}{2}$ to bank 3 and $\frac{1}{2}$ to bank 2, and bank 2 owes $\frac{1}{3}$ to bank 1 and $\frac{2}{3}$ to bank 3. If the initial cash vector is the minimum vector $\mathbf{c}^{m}=\left(\frac{2}{3},\frac{1}{2},-\frac{7}{6}\right),$ then at step 1 bank 1 sends $\frac{1}{2}$ to bank 3 and $\frac{1}{6}$ to bank 2; at step 2, bank 2, having $\frac{1}{2}$ from its initial holding and $\frac{1}{6}$ obtained from bank 1 at step 1, sends $\frac{1}{3}$ to bank 3. All debts are paid, and the final cash position is $\mathbf{c}=(0,0,0).$
\end{example}

We call a structure of obligations and \crsp payments a \emph{cascade} structure, or simply model is a cascade, if set $J$ can be partitioned into sets $(J_1,J_2,...,J_m)$ such that the structure of obligations has the \flw \ form: there is a group of banks $J_1$ without any debts, there is a group of banks $J_2$ that owes money only to banks in $J_1$, etc, and, finally, there is a group of banks $J_m,$ $m\leq n$, to which no bank owes money, and these banks may have debts to all other banks. To simplify presentation in all examples we consider the situation when each set $J_k$ has only one bank, and then $m=n$. The cascade described in Example \ref{ExampleTwo} is depicted in Figure 1(b).

\begin{figure}[t] 
\label{FigureA}
  \centering
  \includegraphics[scale=0.45]{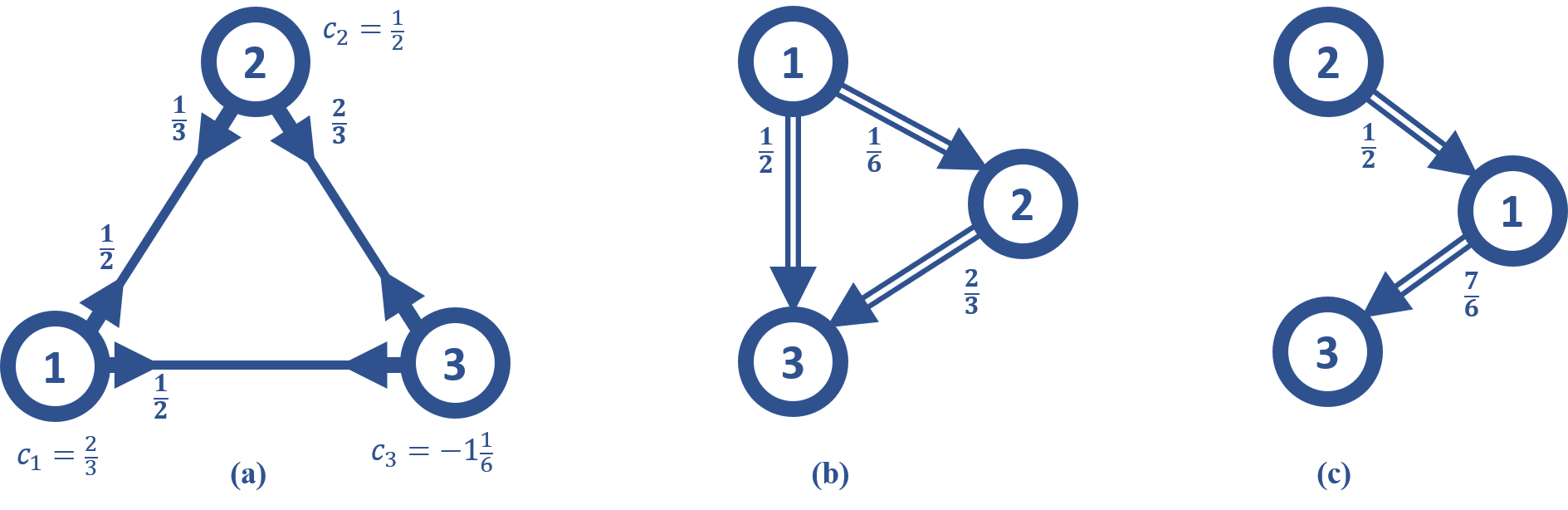} 
   {\small \caption{(a) The matrix of relative liabilities and initial cash vector from Example \ref{ExampleTwo}. (b) Cascade 1: transfers result in full repayment. (c) Cascade 2: an alternative cascade of transfers resulting in full repayment.}}
\end{figure}

Surprisingly, even in this very simple example, the transactions sequence is
not unique -- not only in timing of payments, but also in the order of
states in the cascade. Let us obtain a second cascade using another
mechanism of debt restructuring for the same model and the sufficient cash
vector (see Figure 1(c)). For a sufficient model, the pairwise obligations can be reduced to
unilateral ones: bank 1 owes $\frac{3}{2}=\frac{13}{2}-5$ to bank 3, bank 2 owes $
\frac{5}{6}=\frac{22}{3}-\frac{13}{2}$ to bank 1, and bank 3 owes $\frac{1}{3}=20\frac{3}{4}-22\frac{2}{3}$
to bank 2. Since these obligations form a closed circuit, the amount
$\frac{1}{3}$ can be subtracted and now the obligations are: bank 3, as in the first
cascade, has no debts, banks 1 owes $\frac{7}{6}=\frac{3}{2}-\frac{1}{3}$ to bank 3, and banks 2 owes $\frac{3}{6}=\frac{7}{6}-\frac{1}{3}$ to bank 1. Then the second cascade is at step 1, bank 2, having $\frac{1}{2}$, sends $\frac{1}{2}$ to 1; at step 2 bank 1, having $\frac{2}{3}$, adds $\frac{1}{2}$ obtained at step 1, and sends $\frac{7}{6}$ to 3.

Theorem \ref{Th3} in Section \ref{SectionCascade} establishes that any sufficient system could be modified, via a string of admissible payments, to an equivalent system that has a cascade structure. In Subsection \ref{ssCAlg}, we describe a recursive algorithm to construct the cascade.

\section{Setup}\label{Setup}

In our model, the information is summarized by the triplet $M=(\mathbf{b},Q,%
\mathbf{c})$, where $\mathbf{b}=(b_{1},...,b_{n})$ is the debt vector, the
stochastic matrix $Q=\{q_{ij}\}$ describes relative liabilities, and the
vector of initial cash positions $\mathbf{c}=(c_{1},...,c_{n})$ has the same
meaning as in the original Eisenberg and Noe model with one important distinction: we allow $c_i\leq 0$ for some or even all $i$.\footnote{Sometimes vector $\mathbf{c}$ is interpreted as vector of outside assets \citep{Elsinger2009,glasserman2016}; a negative entry corresponds to outside liabilities.} The matrix of liabilities $B=\{b_{ij}\}$ uniquely determines matrix $Q=\{q_{ij}\}$ and the debt vector $\mathbf{b,}$ and \emph{vice versa}: $%
b_{ij}=b_{i}q_{ij}$ for any $i,j$. By convention, if for some bank $b_{i}=0,$
yet there is $k$ with $q_{ki}>0$, then we define $q_{ii}=1$ and $q_{ij}=0$
for all $j\neq i$.

Since one of our goals is to obtain solution $\mathbf{p}$ of (\ref{Basic}),
we assume that the conditions (A) and (B) are always fulfilled.

In our model, the parameters that describe each bank $i$, $b_{i},c_{i},p_{i}$
depend on time that runs on a finite interval starting with $t=0.$ The
initial position of bank $i$ is $x_{i}(0)=(b_{i},c_{i})$. The position
of bank $i$ at time $t$ is described by the vector $%
x_{i}(t)=(b_{i}(t),c_{i}(t))$, representing its \textit{remaining} debt and
its \emph{current} cash position, and the system a whole is decribed by a $%
2n $-dimensional, vector $\mathbf{x(t)}=(x_{i}(t),i=1,...,n)$. The amount
that bank $i$ paid until moment $t,$ $p_{i}\left( t\right) $ is a function
of these two parameters, $p_{i}(t)=b_{i}-b_{i}(t).$

One can visualize the continuous-time model as follows. Each bank $i$ is a
tank filled by a fluid (money) with initial level $c_{i}=c_{i}(0),$ which
might be positive as well as negative. Each tank is connected to all other
tanks by \emph{inlet} and \emph{outlet} pipes. The \textit{maximum possible
rate} of the flow through each pipe $(i,j),$ which is an outlet for $i$ and
an inlet for $j,$ is its \textit{capacity} $q_{ij}$. Since matrix $Q$ is
stochastic, the total capacity of all outlets from tank $i$ is $1$.
Bank $i$ pays its debts with out-rate $u_{i}(t)$, the intensity with
which money flow from tank $i$ to other tanks. Thus, at any point in time,
bank $i$ pays to bank $j$ with rate $q_{ij}u_{i}(t)$ proportional to its
total debt in compliance with assumption (A).

To fully specify the continuous-time model, we add assumptions (C1) and (C2)
to assumptions (A) and (B).\vspace{0.2cm}

\noindent(\textbf{C1)} \emph{At any moment }$t,$ \emph{all out-rates must satisfy} $%
0\leq u_{i}(t)\leq 1$.\vspace{0.2cm}

The out-rates $u_{i}(t)$ completely specify the in-rates $%
n_{i}(t)=\sum_{j\in J}u_{j}(t)q_{ji}$, which allows us to focus on the
former. Note that since the in-rates $n_{i}$ are defined by the \emph{columns%
} of the stochastic matrix $Q,$ they can potentially be less or more than
one.

To formulate assumption (C2) about in-rates $u_{i}(t)$ for different banks,
we need to introduce two other interrelated notions, which will play a key
role in our analysis. Define a \emph{partition} of the set of all banks $J$
into three groups, $P(t)=(J_{+}(t),J_{0}(t),J_{\ast }(t))$. A\ partition is
a function of time $t;$ that is, a bank belongs to one of the three groups
at each moment $t\geq 0$. The first group of banks $J_{+}(t)$ is called
\textit{positive}, and consists of those banks that, at time $t$, still have
outstanding debt and have positive cash value. The out-rates in this group
can have any values $u_{i}(t),$ $0\leq u_{i}(t)\leq 1$. The next group $J_{0}(t) $, called \textit{zero}, are those banks that have positive debt and zero or even negative cash value. They might be still paying their debts with money flowing to them from other banks. The out-rates in this group must coincide with the in-rates, $u_{i}(t)=n_{i}(t)$ at any moment in time. The last group, $J_{\ast }(t)$, called \textit{paid-up}, includes those banks that have paid
out all of their debts, and are still, possibly, receiving money from other
banks. For them the out-rates are equal to zero, $u_{i}(t)=0.$

Summing up, at each moment $t$ we have a partition of all banks into three
groups
\begin{eqnarray}
J_{+}(t) &=&\{i:b_{i}(t)>0,c_{i}(t)>0\};  \notag \\
J_{0}(t) &=&\{i:b_{i}(t)>0,c_{i}(t)\leq 0\};  \label{Jt} \\
J_{\ast }(t) &=&\{i:b_{i}(t)=0\}.\   \notag
\end{eqnarray}%
Being in one of these groups determines the bank's \emph{status} at moment $t$. 

Assumption (C2) specifies out-rates $u_{i}(t)$ for all banks at all times $%
t\geq 0$:
\begin{eqnarray}
\hspace{-2cm}(\mathbf{C2})\hspace{2.2cm}\text{If}\ \ i &\in &J_{+}(t),\ \text{then}\ \
u_{i}(t)=a_{i},\quad 0<a_{i}\leq 1;  \notag \\
\text{If}\ \ i &\in &J_{\ast }(t),\ \text{then}\ \ u_{i}(t)=0.  \notag\\
\text{If}\ \ i &\in &J_{0}(t),\ \text{then}\
u_{i}(t)=n_{i}(t)=\sum_{j\in
J_{+}(t)}a_{i}q_{ji}+\sum_{j\in J_{0}(t)}u_{j}(t)q_{ji}, \label{int}
\end{eqnarray}

We denote $d_{i}(t)=n_{i}(t)-u_{i}(t)$, the \emph{balance} rate, since this
coefficient defines whether cash position $c_{i}(t)$ will grow, decline, or
remain constant. For the zero group, the out-rates are necessarily equal to the
in-rates, so $d_{i}(t)=0$. Banks with positive and zero status, i.e., banks continuing to pay their debts we call \emph{active}, other banks are nonactive. We say that bank $j$ is a \emph{sponsor} of bank $i$ at moment $t$ if it pays its debt to $i$, i.e., if $p_{ji}(t)>0$.

For any subset $B\subseteq J$ of the set of all banks $J=\{1,2,...,n\}$, $%
Q_{B}$ will denote the matrix obtained from stochastic matrix $Q$ by
deleting all rows and columns not in $B$; $I_{B}[=I]$ denotes the identity
matrix of corresponding dimension. Vector $v_{B}$, made from $n$-dimensional
vector $v$, is defined similarly.

The selection of positive constants $a_i$ for the positive group of banks is a matter of convenience, but we will use only two options. If the set $J_{0}(0)$ is empty, we assume that all $a_i=1.$ We call this a \emph{regular} case and refer to this flow as the \emph{basic flow}. For example, if $c_{i}=c_{i}(0)>0$ for every bank $i,$ we are dealing with a regular case.

The only other values of constants $a_i$ that we are going to use are $a_i=\pi_i>0$ for $i\in B\subseteq J$, where $\pi_i$ are given by the invariant distribution of a corresponding Markov chain if matrix $Q_B$ is ergodic. We call them the \emph{invariant rates}.

If there exists a bank $i$ such that $c_{i}=c_{i}(0)\leq 0,$ the solution to (\ref{Basic}) is not necessarily unique. We deal with this case in section \ref{Swm}. Theorem \ref{Th2} characterizes the parameters for which there is a multiplicity of solutions. In addition, there is also a possibility that for some banks with $c_i=0,$ the in-rates may exceed $1$ and thus some of these banks should be instantly reclassified as positive, since they will have $c_i(t)>0$ on some nonzero interval $(0,\delta)$. We will discuss the so-called Big Bang effect in Subsection \ref{SubsectionBigBang}.\footnote{For a similar reason, the Big Bang cosmology assumes that the life of the Universe
starts after the \textquotedblleft Plank epoch", a minimal period
of time, has passed.}
%%%%%%%%%%%%%%%%%%
\section{The Regular Case} \label{SubsectionRegularCase}

%%%%%%%%%%%%%%%
In this Section, we start with stating formally our Theorem \ref{Main} that focuses on the case when all banks have a positive amount of cash at the initial moment: $c_{i}>0$ for all $i$, i.e., the set of zero banks $J_{0}(0)=\varnothing$, and basic out-rates $u_i=1$ are used for a positive group. Then, we describe how the evolution of the continuous-time model results in an Eisenberg-Noe clearing vector in finite time and characterize the discrete dynamical system that describes the evolution, completing the proof.

\subsection{The Existence Theorem}

What happens with when all pipes are open at time $0$%
? The assumptions (A), (B), (C1), and (C2) jointly determine the following
dynamics:
\begin{eqnarray}
p_{i}(t) &=&\int_{0}^{t}u_{i}(s)ds,\ \ \   \notag \\
b_{i}(t) &=&b_{i}-p_{i}(t),  \notag \\
c_{i}(t)
&=&c_{i}+\sum_{j}p_{j}(t)q_{ji}-p_{i}(t)=c_{i}+\int_{0}^{t}d_{i}(s)ds,\
\label{PBC} \\
d_{i}(s) &=&n_{i}(s)-u_{i}(s).  \notag
\end{eqnarray}

The last equation determines the balance rate $%
d_{i}(s)$, where $n_{i}(s)=\sum_{j}u_{j}(s)q_{ji}$ is the
in-rate for bank $i$ at moment $s.$

The money will flow exactly until there is at least one bank with a positive
status, i.e., having an outstanding debt and a positive cash reserve. Before that moment, the total amount of debt in the system is monotonically decreasing with at least the unit rate, and thus, since the total amount of debt is finite,
the process ends in a finite time $T_{\ast }=\min \left\{
t:J_{+}(t)=\varnothing \right\} $. Theorem \ref{Main} describes the main characteristics of this process for a regular initial vector $\mathbf{c}$.

\begin{theorem}
\label{Main} (a) For any regular initial vector $\mathbf{c},$ there exists a finite time $T_{\ast }=\min \left\{ t:J_{+}(t,\mathbf{c})=\varnothing \right\},$ at which all out-rates $u_{i}(T_{\ast })=0$ for every bank $i.$ The time interval $[0,T_{\ast })$ consists of a finite number $k_{\ast }$
of half-intervals $\Delta_k=[T_{k},T_{k+1}),$ $k=1,2,...,k_{\ast }$, $T_{1}=0,$ $%
T_{k+1}=T_{\ast }$. Out-rates $u_{i}(t)$ are constant on these intervals; functions $p_{i}(t),$ $b_{i}(t),$ $c_{i}(t),$ $0\leq t\leq T_{\ast }$, satisfying equations (\ref{PBC}) are correspondingly linear on each interval $\Delta_k$. The moments $T_{k}>0 $ are the moments when at least one bank changes its status, i.e., one of the functions $b_{i}(t),$ $c_{i}(t)$ hits zero from above. Vector $\mathbf{p}(T_{\ast })$ solves equation (\ref{Basic}), i.e., it is
the clearing payment vector in both the Eisenberg-Noe and continuous-time
models.

(b) The following \emph{Monotonicity Property} holds: functions $u_{i}(t)$ are nonincreasing in $k$ for all $i\in J$, i.e., $u_i\left(T_k\right)\geq u_i\left(T_{k+1}\right),\ k=1,2,...,k_{\ast }-1.$

(c) Partition $P_{k}=P(T_{k})=(J_{+,k},J_{0,k},J_{\ast ,k}),$ and vectors $%
\mathbf{b}_{k}=(b_{i}(T_{k}))$ and $\mathbf{c}_{k}=(c_{i}(T_{k}))$ contain
all the information about the system at moment $T_{k}$ and uniquely define
out-rates $u_{i,k}$ (and thus all other functions) on the next interval $\Delta_k$.
\end{theorem}

Because the evolution of the system is interesting in itself, we will prove parts (a) and (c) of Theorem \ref{Main} in the next subsection,  and delegate a more technical proof of part (b) to Appendix.

\subsection{The Evolution of Dynamical System}\label{ssEDS}

When the process ends at $T_{\ast }=\min\left\{ t:J_{+}(t)=\varnothing \right\},$ we have the following picture. As the first of the balance conditions (\ref{ge}) shows, the sum of all cash positions is constant at all times. Thus, at the final moment $T_*$, we have $$\sum_ic_i=\sum_ic_i(T_*)=\sum_{i\in J_*(T_*)}c_i(T_*).$$ The last equality holds because $J_+(T_*)=\varnothing$ and $c_i(T_*)=0$ for all $i\in J_0(T_*)$. These equalities immediately imply that the set $J_*(T_*)$ is always non-empty even if all $c_i\ll b_i$, and the sum is positive.

What happens before this final moment? According to Assumption (C2) all out-rates, and hence in-rates as well, are constant on any interval where statuses are not changed, and therefore all cash (positions), debts and payments are linear functions on these intervals. Thus, in the regular case, when all $c_{i}(0)>0$, all banks have positive status until some moment $T_1$ when either one debt or cash level hits zero. Since all debts go down with a unit speed, $T_1<\infty$. Without loss of generality, perturbing slightly, if necessary, the initial conditions, we can assume that at the moments of status change only one bank changes its status.

At any moment $t,$ the status change may occur only when some positive bank
paid up its debt, or its cash position hits zero, or some
zero bank paid up its debt. In the first case, the zero group remains the
same, and in-rates for all banks, including the solution of (\ref{int}) for
zero group, did not increase. The equality $c_{i}(t)=c_i$ holds for each bank in zero group. In the second case, when one positive bank changes its status to zero, this is possible only when the balance rate $d_{i}$ of this banks was negative, i.e., rate in $n_{i}$ was less than the rate out $u_{i}=1$. Then this bank is added to zero group, and the rate out from this bank
drops from $u_{i}=1$ to the new value $n_{i}<1$, defined by a new
solution of (\ref{int}). The in-rates for other banks in zero group are also
decreased since the total contribution to this group from positive sponsors
decreased. In the third case, the situation is similar.

Thus, the incoming rates are always non-increasing for all banks. The formal proof that in-rates in zero group, and in all groups, no matter if this group is expanding or diminishing, are non-increasing, i.e., point (b) of Theorem \ref{Main}, is given in Appendix.

Assuming that point (b) of Theorem \ref{Main} holds, the interval $[0,T_{\ast })$ consists of a finite number $k_{\ast }$
of half-intervals $\Delta _{k}=[T_{k},T_{k+1}),k=1,2,...,k_{\ast },T_{0}=0$,
where the times $T_{k}>0$ are the only times when one bank changes its status, i.e., one of the functions $b_{i}(t),c_{i}(t)$ hit zero, being positive before. Plus, (b) yields an important \emph{irreversibility property} for a status change after initial moment: The only possible status change for all $t>0$ is: from the positive group to the paid-up group, or to the zero group, and from the zero group to the paid-up group. Because of
the irreversibility, the number of such moments is no more than $2n$. For sake of brevity, we call these intervals $\Delta_k$ \emph{status intervals}.

Note that in general situation when at moment $t=0$ there are banks with $c_i\leq 0$ an instant status change is possible. This is due only to possible reclassification of some zero banks to positive (see the Bang Bang effect in Subsection \ref{SubsectionBigBang}).

To complete the proof of (a) in Theorem \ref{Main}, we will need the following slightly less straightforward result.
\begin{lemma}
\label{P4} The amounts paid by each bank at moment $T_{\ast }$,
i.e., $p_i=p_{i}(T_{\ast }),$ $i\in J$, gives a solution to the clearing equation (\ref{Basic}).
\end{lemma}
\begin{proof} The status of each bank $i$ at moment $T_{\ast }$ is either zero or paid-up. If it is paid-up, then the total payment of this bank coincides with its debt, i.e., $p_{i}(T_{\ast })=b_{i}$. If bank $i$ at $T_{\ast }$ has zero status, i.e. $c_{i}(T_{\ast })=0$, the debt $b_{i}$ is not fully paid, $p_{i}(T_{\ast })< b_{i}$. Since the rate of payment of each banks $j$ to bank $i$ at each time $t$ is proportional to capacity $q_{ji}$, the total payment obtained by bank $i$ from bank $j$ is $p_j(T_{\ast})q_{ji}$. Using the equality $c_{i}(T_{\ast })=0$  and the second of balance equalities (\ref{ge}) we obtain that the total payment of bank $i$, is equal to the sum of the initial cash $c_i$ and the money obtained from other banks, i.e., $p_{i}(T_{\ast })=c_{i}+\sum_{j}p_{j}(T_{\ast })q_{ji}<b_i$. Thus, in both cases the amount paid by each bank satisfies clearing equation.
\end{proof}

\subsection{Discrete Dynamics}\label{ssDDS}

Our next goal is to describe all information needed to compute the end time of each status interval $\Delta_k$ and complete the proof of part (c) of Theorem \ref{Main}.

According to our definitions and part (a) of Theorem \ref{Main}, we have continuous flow $X(t)$ defined on the interval $[0, T_*)$ by vector $X_{}(t)=(t,P_{}(t),\ b_{i}(t),\ c_{i}(t),i\in J)$. On each status interval, all out-rates are constant and define a simple linear dynamics for all parameters of the model.  As a result, we can utilize one of the main computational advantages of our model: its dynamics is fully characterized by a homogeneous \emph{discrete dynamical system} $(X_{k})$, where time moments $k=1,...,k_{\ast }+1$ correspond to moments $T_{k}$ in the continuous-time model, and vector $X_{k}$ contains all information about the system at this moment. 
This vector is: $X_{k}=(T_{k},P_{k},\ b_{i,k}=b_{i}(T_{k}),\ c_{i,k}=c_{i}(T_{k}),i\in J)$, where $P_{k}=P(T_{k})$ is a partition of a system at the beginning of the next interval $\Delta _{k}=[T_{k},T_{k+1})$.

When it is clear which interval $\Delta_k$ is considered, to simplify notation, we suppress further indices $k$, denoting $\Delta _{k}=\Delta $,\ $T_{k}=T$,\ $T_{k+1}=T^{\prime }$, out-rates $u_{i,k}=u_{i}$, etc.
The values of out-rates for the basic flow are selected at moment $T_{k}$ for the next interval $\Delta_k$ according to (C2) and partition  $P_{k}=P(T_{k})=(J_{+},J_{0},J_{\ast })$, i.e.:
$u_{i}=1$ if $i\in J_{+}$, $u_{i}=0$ if $i\in J_{\ast}$ and if $i\in J_{0}$, then $u_{i}=v_{i}$, where vector $v=(v_i)$ is obtained as a solution of a linear non-homogeneous system of equations (\ref{int}). Thus, we have
%%%%%%%%%%%%% (8)         n_{i}=\sum_{j\in J_{+}}q_{ji}+\sum_{j\in J_{0}}u_{j}q_{ji}.
\begin{eqnarray}
b_{i}(t)&=&b_i(T_{k-1})-u_{i}t, \notag \\
c_{i}(t)&=&c_{i}(T_{k-1})+(n_i-u_{i})t.   \label{pbc}
\end{eqnarray}
%%%%%%%%%%%p 18 ln 616 pbc = 9        {k-1}
The time of the next status change $T^{\prime }=T_{k+1}$ is the first time when one of those quantities $b_{i}(t),c_{i}(t)$, which were positive at moment \ $T=T_{k}$, will reach zero for the first time. Assuming that out-rates on interval $(T,T+t)$ are equal to $u_{i}$, we denote the \emph{potential} moment $T+s_{i}$ when bank $i$ will repay its debt, and the \emph{potential} moment $T+t_{i}$ when bank $i$ reaches the zero cash position. Equations (\ref{pbc}) immediately imply that
$$s_i =\begin{cases} b_{i}(T)/u_{i}, & \text{if}\ \ b_i(T)>0, \ \text{and}\ \ u_i>0; \\
                    +\infty & \text{if}\ u_i=0.
       \end{cases}$$
$$t_i =\begin{cases} -c_{i}(T)/d_{i}, & \text{if}\ \ c_i>0,\ \text{and}\ \ d_{i}<0; \\
                    +\infty & \text{if}\ c_i=0 \ \text{or}\ d_{i}\geq 0.
       \end{cases}$$
Finally,
\begin{equation}
\label{T}
T^{\prime }=T+\min_{i}(s_{i},t_{i}).
\end{equation}
 Note that for $i\in J_{0}$ we have $c_{i}(t)\equiv c_i\leq 0$ for all $t\in (T,T')$, so for them $t_i=+\infty $.\par
%%%%%%%%%%%%%%%
Thus, the state $X_{k}$ with out-rates specified by (C2), (\ref{int}), and a simple system of linear  equations uniquely determines the moment of the next status change $T_{k+1}$, and therefore all the values $X_{k+1}$. Thus we proved point (c) of Theorem \ref{Main}. We call this transformation of $X_k$ into $X_{k+1}$ mapping $G.$ Note that this mapping does not depend on $k$, which makes it easily programmable. 

The following example illustrates the dynamic evolution of the system.
%%%%%%%%%%%%%%%%%%
\begin{example} 
\label{ExampleThree}
Let $n=3$,  matrix $Q$ be given by (\ref{MatrixQ}), and the same debt vector $\mathbf{b}=(13, 22, 20)$ as in Example \ref{ExampleTwo}, but with cash vector $\mathbf{c}=\left(\frac{1}{2},\frac{1}{2},0\right)$. For this vector $\mathbf{b},$ formula (\ref{cm}) yields that $\mathbf{c}^{m}=\left(\frac{2}{3},\frac{1}{2},-\frac{7}{6}\right)$, i.e., $\mathbf{c}\prec \mathbf{c}^{m}$, so this cash vector makes model deficient. Formally, this is not a regular case, since $J_0(0)=\{3\}$, but this is an example of the Big Bang effect that can be easily fixed, because $n_3=\frac{1}{2}+\frac{1}{3}>1$, and thus bank 3 should be reclassified as positive. 

Thus, on the first interval $\Delta_1,$ we have \emph{all} $u_i=1$. Then $\mathbf{d}=\left(-\frac{5}{12},\frac{1}{4},\frac{1}{6}\right).$ Each moment of status change $T_i=T_{i-1}+t_i, i=1,2,..$, is the moment when either some debt is paid up or some cash position hits zero from above. The moment $t_1=T_1$ is defined by condition $c_1(t)=c_1-d_1t=0$, which yields $t_1=\frac{6}{5}$. Then $\mathbf{c}(T_1)=\left(0,\frac{4}{5},\frac{1}{5}\right)$ and $\mathbf{b}(T_1)=(b_i-u_it_1)_i=\left(11\frac{4}{5},20\frac{4}{5},18\frac{4}{5}\right).$

On the second interval $\Delta_2,$ we have $J_+=\{2,3\}$ and $J_0=\{1\};$ thus, $u_2=u_3=1$ and $u_1=n_1=\frac{7}{12}$. Then
$\mathbf{d}=\left(0, \frac{1}{24},-\frac{1}{24}\right).$  Moment $t_2$ is defined by condition $c_3(t)=\frac{1}{5}-d_3t=0$, and then $t_2=4\frac{4}{5},$ $T_2=t_1+t_2$. Then $\mathbf{c}(T_2)=\left(0,\frac{4}{5}+d_2t_2=1,0\right)$ and $\mathbf{b}(T_2)=(b_i(T_1)-u_it_2)=(9,16,14)$.

On the third interval, $\Delta_3,$ we have $J_+=\{3\}$ and $J_0=\{1,2\},$ which yields $u_2=1$ and since $c_1(T_2)=c_3(T_2)=0$, we need to solve the \eqm system (\ref{int}) with two equations $u_1=\frac{1}{3}+\frac{1}{4}u_3,$ $ u_3=\frac{2}{3}+\frac{1}{2}u_1$  to find $u_1=n_1$ and $u_3=n_3$. This system has a solution $u_1=\frac{4}{7},$ $u_3=\frac{20}{21}$. Then $n_2=1$ and $\mathbf{d}=(0,0,0)$. Thus, $\mathbf{c}(t)=\mathbf{c}(T_2)=(0,1,0)$ and moment $t_3$ is defined by the moment when one of \emph{debts} $\left(9-\frac{4}{7}t_3,16-t_{3},14-\frac{20}{21}t_{3}\right)$ hits zero. Then $t_3=14\frac{7}{10}$ with $b_3(T_3)=0$. At moment $T_3=t_1+t_2+t_3,$ we have $\mathbf{b}(T_3)=\left(\frac{3}{5},1\frac{3}{10},0\right)$ and $\mathbf{c}(T_3)=(0, 1, 0)$. 

On the last interval $\Delta_4,$ $J_0=\{1\},$ $J_+=\{2\},$ and $J_*=\{3\}$, and thus 
 $u_1=n_1=\frac{1}{4},$ $u_2=1,$ and $ u_3=0$. Then $d=\left(0,-\frac{5}{6},\frac{5}{6}\right).$  Moment $t_4$ is defined on this interval by condition $c_2(t)=1-d_2t=0$, and then $t_4=1\frac{1}{5}$. At moment $T_4=T_*=t_1+t_2+t_4+t_4,$ the flow stops because bank 2 was the last positive bank. Then cash vector $\mathbf{c}(T_*)=(0, 0, 1)$ and $\mathbf{b}(T_*)=b_i(T_3)-u_it_4)=\left(\frac{1}{5},\frac{1}{10},0\right),$ with banks 1 and 2 in default.
\end{example}

\section{ \textquotedblleft Swamps\textquotedblright\ and Multiplicity of
Solutions}\label{Swm}

In the regular case (all $c_{i}>0$ and $u_{i}=1$ for positive group) there is a unique solution to the dynamical system as described in Theorem \ref{Main}. In this section, we will formulate the results for the general case when there are banks with initial cash position at zero or even negative. If banks from positive group have some liabilities to these banks, then after a possible reclassification, all remaining banks in the zero group have in-rates, obtained as a solution of system (\ref{int}), less than one and therefore they can be treated as zero group in regular case and nothing new happens. The new interesting case is when there are banks in zero group such that banks with positive cash have no obligations, direct on indirect, towards them. This case might look exceptional, yet it is critical to understanding of potential multiplicity of clearing vectors. As we will see, the main reason for multiplicity is that some debts can be settled between banks that do not have any money.

Informally, a \emph{swamp} is a set, where all its members at $t=0$ have no positive cash, no direct or indirect inflow from positive group, and have positive debts but only to themselves. As a result, they do not participate in the basic flow, but they have a possibility to ``run money in a circle", paying partially their debts, at the same time keeping their nonpositive cash positions. Such transactions can be considered as a partial debts restructuring, i.e., a transformation of the initial debt vector into a new one. We will give a formal definition of a swamp in Subsection \ref{SubsectionSwamps}

\subsection{The Multiplicity Theorem}\label{ssTh2}
We start by using the matrix of initial obligations $Q=Q(0)$ to divide all banks that are not in the paid-up group $J_{*}(0)$ into two disjoint sets. Denote $A_+$ the set of all banks that at moment $t=0$ are either positive, or have sponsors from positive group, or have sponsors from the previous group, etc., i.e., have positive direct or indirect sponsors. Formally, $A_{+}=J_+(0)\cup J_{+,1}\cup J_{+,2}...$, where $J_{+,1}$ is a set of zero banks that have sponsors from $J_{+}(0)$, $J_{+,2}$ is a set of zero banks that have sponsors from $J_{+,1}$, etc. We define all
other banks that are not in the paid-up group as $A_{0}$. These are banks that have no positive cash and no positive input from the outside at the initial moment and hence forever. They may have liabilities to banks in $A_{+}$ and $J_{*}(0)$ but not \emph{vice versa}. The dynamics of a flow in set $A_+$, aside for the possible reclassification of some zero banks at $t=0$ due to the ``Big Bang'' effect, which we discuss in Subsection \ref{SubsectionBigBang}, is unique and is described in Theorem \ref{Main}. The evolution of group $A_0$ is the main  subject of Theorem \ref{Th2}.

Importantly, swamps may exist at the initial moment, but they can not appear later (Lemma \ref{L5Evol}). In model where the assumption of ``no debt seniority'' might be modified, swamps could appear in later stages as well. Of course, they do appear in reality, as we mentioned in Introduction the case of \emph{Enron} and more recent debacle with \emph{Wirecard} in Germany.

Now we can formulate Theorem \ref{Th2} which extends the results of Theorem \ref{Main} to general flows. Let us represent set $A_0$ as a union of disjoint sets $A_0=J_{0,*}+U_0+S_0$, where $J_{0,*}$ is the set of all paid-up banks in $A_0$, $U_0$ is the set of all transient banks in $A_0$, and $S_0$ is the union of all ergodic subclasses in $A_0$. (Formal definitions are given in the next Subsection \ref{ssMC}; for now, it is sufficient to say that banks from $S_0$ will be the ones that can pay each other without any cash on hands.)

\begin{theorem}
\label{Th2}
(a) For any initial cash vector $\mathbf{c}$, there exists a basic flow solution of (\ref{Basic}), $\mathbf{p},$ with $p_i=0$ for $i\in A_0$.

(b) If set $S_0$ in the decomposition of $A_0$ is empty, then the basic flow solution is a unique solution of (\ref{Basic}), and all banks in $A_0$ are nonactive.

(c) If set $S_0\neq\emptyset$, i.e., if there is at least one swamp, then there are multiple solutions $\mathbf{p'}$ of the form $\mathbf{p'}=\mathbf{p}+\sum_{k}s_k\mathbf{p_{*k}}$, where $\mathbf{p}$ is the basic flow solution, $0\leq s_k\leq 1$, and each vector $\mathbf{p_{*k}}$ is obtained through an invariant \dst \ $\pi_{*k}$ for the \crs ergodic subclass.\end{theorem}

Note that our basic flow solution corresponds to the \emph{smallest}, and solution with all $s_k=1$ in Theorem \ref{Th2} to the \emph{greatest} clearing payment vector in \cite{eisenberg2001systemic} and \cite{kabanov2018clearing}. %\cite{KaMoBi17}.
They coincide if there are no swamps.

The existence of swamps makes it possible that the flow in swamp(s) will last more that $T_*$, which is defined as a moment when the last \emph{positive} bank repays its debt.

\subsection{Markov Chains, Basic Flow Evolution, and Invariant Rates}\label{ssMC}

Analysis of ``swamps'' and proof of Theorem \ref{Th2}  requires introduction of a new machinery in the studies of financial clearing, that of Markov chains.

So far, we have not used any stochastic interpretations in our analysis of our deterministic model. Still, it is well known that many statements in probability theory have their deterministic analogs and \emph{vice versa}.  The invariant distribution for Markov chain was in fact introduced by Gustav Kirchhoff for electrical networks long time before even the concept of Markov chain was formulated.

\cite{kemeny1976markov} is a classic treatise on finite Markov chains; we refer to \cite{shiryaev2019probability} as an excellent introduction to the subject. If Markov chain $Z=(Z(n))$ has state space $S$ and transition (stochastic) matrix $Q$, and $\mathbf{m_k}=(m_k(i), i\in S)$ is a vector of probability distribution of $Z$ at some moment $k$, then $\mathbf{m_{k+1}}=Q^T\mathbf{m_k}$ is the distribution vector at the next moment. Given any stochastic matrix $Q,$ all states can be divided into \emph{transient} states and \emph{ergodic} states. Each ergodic state is a member of some ergodic subclass. All transient states form a transient set. A Markov chain eventually leaves transient states moving to one of these subclasses to stay there forever. If subset $B$ is such an ergodic subclass, the Markov chain can go from every state to every state (not necessarily in one move). Some textbooks call such subsets \emph{irreducible} and call ergodic if it is also aperiodic. In both cases, there is an invariant distribution $\mathbf{\pi}$ that satisfies the equality $\pi=Q_B^T\pi$. In latter case, this distribution is also the limit distribution, and in particular $Q_B^n\rightarrow A$, where $A$ is a \sth matrix with identical rows equal to $\pi$.

The evolution of stochastic matrix $Q(t)$ that starts with matrix $Q$ at moment $t=0$ reflects the  partition $P(t)$ as follows. If bank $i$ is paid-up, we consider $i$ as absorbing state, i.e., $q_{ii}(t)=1$. For our goals, it is more convenient to distinguish the absorbing and ergodic states assuming that every ergodic subclass has at least two states.

Obviously, matrices $Q(t)$ are constants on the status intervals and change only at moments $T_k$. The set of absorbing states at moment $t$ is just $J_{*}(t)$ but banks in transient set and ergodic subclasses may have positive or zero status. Thus each bank at each moment had a double classification: by status and state.

Now we need to formulate a few facts about transient sets and ergodic subclasses. Suppose that $B\subseteq J$ and states in $B$ are transient. Then $Q_B$ is a substochastic matrix, i.e., the sum of coefficients in some rows is less that one, and the Markov chain leaves set $B$ at some moment with probability $1$ or, equivalently, matrix $Q_B^n$ tends to zero when $n$ approaches infinity. Thus, the matrix $N_B=(I_{}-Q_{B})^{-1}$ is well-defined. Matrix $N_B$  is a called the \emph{fundamental} matrix for substochastic matrix $Q_{B}$, and has a natural probabilistic interpretation. Namely, $n_B(x,y)$ equals to the expected number of visits to state $y$ for a Markov chain with the initial point in $x$ before exit out of set $B$.
We also have $N_B =\sum_{n=0}^{\infty }Q^{n}_B$ and $N_B^T=(I_{}-Q_{B}^T)^{-1}$.

Let $B=J_0(t)$ for some time $t$, and $Q_{B}$ to be the corresponding (sub)stochastic matrix. Let us introduce vector $\mathbf{e_{B}}$, the ``input vector" from ``sponsors"\ of positive group with their out-rates $u_i=1$ to zero group $B.$ The coordinates of $\mathbf{e_{B}}$ are defined by $e_{i,B}=\sum_{j\in J_{+}}q_{ji},i\in B$.  The out-rates for zero group $B$, which are equal to their in-rates, are denoted $v_B\equiv v$. According to (C2) these rates, defined by formula (\ref{int}), satisfy the equation
\begin{equation}
\mathbf{v_B}=\mathbf{e_{B}}+Q_{B}^{T}\mathbf{v_B},  \label{EE}
\end{equation}
The question is when this equation has a solution. The answer is given by the following lemma.
\begin{lemma}
\label{Lemma51}
 (a) If $Q_{B}$ is a \emph{substochastic} matrix then the solution of (\ref{EE}) is given by formula
\begin{equation}
 \mathbf{v_B}=(I_{}-Q_{B}^T)^{-1}\mathbf{e_{B}}\equiv N_B^T\mathbf{e_{B}},  \label{LS}
\end{equation}

(b) if $Q_{B}$ is  a \emph{stochastic} matrix, and $\mathbf{e_{B}}\neq 0$, then the equation (\ref{EE}) has no solution; if $\mathbf{e_{B}}=0$, then all solutions of (\ref{EE}) are proportional to the solution given by the invariant \dst for \crsp Markov chain, $\pi=Q_B^T\pi$.

(c) Two solutions $\mathbf{v}$ and $\mathbf{v^{\prime }}$ of equations $\mathbf{v}=\mathbf{e}+Q^T_B\mathbf{v}$, and $\mathbf{v'}=\mathbf{e'}+Q^T_B\mathbf{v'}$,
 satisfy $\mathbf{v}\leq \mathbf{v^{\prime }}$ if $\mathbf{e}\leq \mathbf{e^{\prime }}$.
\end{lemma}
%%%%%%%%%% (e.g., \citealp{kemeny1976markov})
The proof of points (a) and (b) of Lemma \ref{Lemma51} follows from simple facts from linear algebra. Point (c) follows immediately from point (a), since the elements of matrix $ N_B^T$ are nonnegative. \par

Before we proceed to the next step, we need to describe how invariant out-rates can be used. Suppose that a subset of banks $B$, say at moment $0$, forms an ergodic subclass with the invariant \dst $\pi$. According to Lemma \ref{Lemma51} and partition of all banks into sets $A_+$ and $A_0$ (see subsection \ref{ssTh2}), if set $B\subseteq A_+$ then it may have positive and zero group members, if set $B\subseteq A_0$ then it has only zero group banks. In both cases we can use the \nvt \prbs $\pi$ as out-rates $u_i$. By the definition of $\pi$ this guarantees that out-rates are equal to in-rates for each bank in $B$. This implies that all $c_i(t)$ remain equal to the initial values $c_i$ till moment when these out-rates stopped to be applied. At the same time each $b_i(t)=b_i-tu_i$ goes down with positive rate $u_i$. Let $t_i=\max_t\{t:b_i(t)>0\},$ $T=\min_{i}t_{i},$ and $i_1$ be the bank where this minimum is achieved. As usual, we assume that there only one such bank, say $j$. It means that $b_j(T)=0$, this bank has no \lbls to all banks in $B$, and since $B$ is a closed subset, no \lbls to all banks in $J$, i.e., its status becomes paid off. For all other banks in $B$, $b_i(T)=b_i-Tu_i=b_i^{(1)}>0$. Thus we restructured debts and our model without changing cash positions of all banks in $B$.

Note that at moment $T$ all banks in $B\setminus j$ become transient because some of them definitely have obligations to bank $j$ because before this moment by definition of an ergodic set they were connected to bank $j$. Therefore, they will remain transient forever. We denote the \crsp payment vector obtained in this process as $p^*_B$. Note that we can change out-rates $u_i=\pi_i$ to some proportional $s\pi_i$ with some coefficient $s>0$. Since $\max_i\pi_i=q<1$, we can select any $0<s<1/q$ without violation of our assumption (C1) that guarantees that out-rates cannot exceed $1$. Then the time of restructuring $T$ will be also changed proportionally to $sT,$ but the payment vector will be the same because the rate of payment will be increased (decreased) in proportion to decreased (increased) period of payment, see equations (\ref{pbc}). Of course if payments will be stopped earlier then one of the debts hit zero, clearing vector will be smaller.

The important point about this construction is that the cash position of each bank in set $B$ was not changed, so it was possible that some of these banks did not have cash at all or even had negative positions. In particular, we obtained the following result.
\begin{lemma}
\label{L53} Let set $B$ be an ergodic subclass and all $c_i\leq 0, i\in B$. All nontrivial solutions of clearing equation (\ref{Basic}) that are limited to this set can be obtained using only invariant out-rates. All these solutions are proportional to vector $\mathbf{p}^*_B$, with the coefficient of proportionality not exceeding $1$.
\end{lemma}
We illustrate this construction by a simple example.

\begin{example}
Let $n=3$, $\mathbf{b}=(2,3,4)$, $c_i\leq 0,$ $i=1,2,3$, and matrix $Q$ given by (\ref{MatrixQ}). It is easy to check that the invariant \dst for the \crsp Markov chain is $\pi=\frac{1}{53}\left(12,21,20\right)$. Let
us select these fractions as the out-rates $u_i$ for these three banks. By definition of $\pi$ we have $u_i=n_i,$ $i=1,2,3$, and therefore $c_i(t)=c_i, i=1,2,3$ for all $t>0$. Then on some interval $\Delta=[0,t)$ we have $b_i(t)=b_i-u_it, i=1,2,3$. Then at moment $T=7\frac{4}{7},$ we obtain that $b_2(T)=b_2-u_2T=0$, and $b_1(T)=b_1-u_1T=\frac{2}{7}>0$, $b_3(T)=b_3-u_3T=\frac{8}{7}>0$. We obtained the clearing vector $p=p(T)=\left(\frac{12}{7},3,\frac{20}{7}\right)$ and the vector of unpaid debts $\left(\frac{2}{7},0,\frac{8}{7}\right)$. The remaining debts of banks 1 and 3 to bank 2,  $\frac{1}{7}$ and $\frac{6}{7},$ respectively, are not repayable, but their \emph{mutual remaining debts} ($\frac{1}{7}$, and $\frac{2}{7},$ respectively) can be cancelled. Ultimately, bank 2 owes to bank 3 nothing and bank 3 owes bank 2 the amount of $\frac{1}{7}$. The model cannot be restructured any further.
\end{example}

\subsection{Swamps} \label{SubsectionSwamps}

The situation when all banks in ergodic set $B$ have zero cash gives rise to an \mpt notion of swamps that in general case lead to a non-uniqueness of solution of a clearing equation. Formally, we say that a \emph{swamp} $B$ is an ergodic subclass of banks that have zero or negative cash at  $t=0.$
% \newpage
The next Lemma \ref{L5Evol} describes the evolution of the banks from the point of view of their positions as states of Markov chains. The irreversibility of a status change is accompanied by the irreversibility of states change. (As above, we assume that only one bank may change its status at a time.) Point (c) establishes formally that swamps might exist at the initial moment, but cannot appear later. Recall that $A_{+}$ are banks that participate in basic flow (regular case), and $A_{0}$ are the banks that have no positive cash and no positive input from the outside at the initial moment and hence forever. They may have liabilities to banks in $A_{+}$ and $J_{*}(0),$ but not \emph{vice versa}.

\begin{lemma}
\label{L5Evol} The only possible evolution of banks as the states of the Markov chain is the following:

(a) paid-up banks are always absorbing, and transition to this state is irreversible;

(b) positive banks, as well as zero banks in $A_{+},$ may be ergodic or transient; as ergodic they may become \trt or absorbing, as transient they may become only absorbing;

(c) zero banks in $A_{0}$ may be transient or ergodic; if transient, they are always transient and inactive, i.e., for them $u_i(t)\equiv 0$ for all $t$; if there is an ergodic subclass, then its members can be active only under invariant out-rates till moment when one of them becomes absorbing, and then all other become transient and inactive. Thus, swamps may exist from the beginning, but cannot appear later.
\end{lemma}

\begin{proof} Part (a) is straightforward. Positive banks maybe ergodic or transient. With a status of an ergodic or a transient bank changed to paid-up, its state becomes absorbing. All other states in the ergodic subset become transient because now they are connected to an absorbing state. If zero bank in $A_{+}$ is transient, its out-rate is positive and hence it may become an absorbing state or remain transient till the end, and it means this bank ends in default. This proves (b).

If $B$ is the set of all transient banks from $A_0$ then the input vector $e_B=0$ for all moments $t$ and by part (a) of Lemma \ref{Lemma51} the only solution for \eqm rates is zero so they nonactive forever.

If $B$ is an ergodic subclass in $A_0$, then the only possible out-rates $a_i, i\in B$, by Lemma \ref{L53}, must be proportional to the invariant out-rates $\pi_i$, where $\pi_i$ is the \nvt \dst for $Q_B$. Only these out-rates can keep $c_i(t)\equiv c_i$ on some interval.  When the invariant out-rates $\pi_i$ are applied, then on some interval
$[0, T)$, all debts will be decreased to the level $b_i-\pi_iT\geq 0$ with exactly one bank $j$ with $b_j(T)=0$ and all other $b_i(T)>0$. At moment $T$ bank $j$ changes its status to paid-up, and its state becomes absorbing. Then all
other banks in $B$ become transient and become nonactive. Thus all transient banks $A_0$ never be able to repay their debts, but banks in each of ergodic subclasses will be able to restructure and diminish their debts to each other. This proves (c).
\end{proof}

Now we have everything to complete the proof of Theorem \ref{Th2}.

\paragraph{Proof of Theorem \ref{Th2}.} Part (a) follows from Theorem \ref{Main} and the fact that $p_i=0$ for $i\in A_0$ obviously satisfies clearing equation (\ref{Basic}).

Part (b) follows from part (a) of Lemma \ref{L5Evol}.

Part (c) follows from Lemmas \ref{L5Evol} and \ref{L53} and the fact that if the number $k$ of the irreducible subclasses is more than one, they have an empty intersection, and therefore any linear combination of these solutions with coefficients $0\leq s_k\leq 1$ is also a solution of (\ref{Basic}) with coordinates $p_{*i}\leq b_i$ for $i\in B$ and $p_{*i}=0$ for $i\notin B.$ $\qedsymbol$

\section{The Cascade Theorem and Cascade Algorithm}\label{SectionCascade}
Introducing the model, we asked the following question: What is the smallest initial cash vector $\mathbf{c}$ such that \emph{all banks} ultimately pay their debts? The answer was the vector $\mathbf{c^{m}}$ satisfying (\ref{cm}). The next natural question is as follows. Given matrix $Q$ and vector $b$, what is the smallest vector $\mathbf{c}$ such that \emph{at least one bank} will pay its debt? The answer to this question might seem surprising: a bank that fully repays its debt exists for \emph{any} initial cash vector $\mathbf{c}=(c_1,...,c_n)$, even if all $c_i<0$.

In this section, we will show that any sufficient model $M =(\mathbf{b},Q,\mathbf{c})$ can be restructured into an equivalent cascade model, in which there is a group of banks that can paid off their debts even with negative cash positions, a group of banks that need to pay only to the banks in the first group, and so on.

\subsection{The Cascade Theorem} \label{ssTh3}
Given any model $M =(\mathbf{b},Q,\mathbf{c})$, we say that payments $\mathbf{p}=(p_1,...,p_n)$ are \emph{admissible} if they do not exceed the obligations of banks and are proportional to their obligations. We say that model $M'=(\mathbf{b'},Q',\mathbf{c'})$ is \emph{obtained from model $M=(\mathbf{b},Q,\mathbf{c})$ by restructuring} if the second model obtained from the first via admissible payments.

Note that we do not assume that these proportions are equal. Under admissible payments, one bank can pay 10\% of its debt and the other 20\%. Also, observe that the clearing vector $p^*$ in the initial model is in fact the maximum possible vector of admissible payments, i.e., $p^*\succ p$ for any admissible $p$.

This definition implies in particular that $b_{ij}\geq b'_{ij}$ for all $i,j$. We also say that $M$ is \emph{reduced }to $M'$, and denote $M \rightarrow M'$. It is easy to see that if $(p^{*})'$ is the clearing vector in this new model, then clearing vector in the initial model $p^*=p+(p^{*})'$. We also may consider a sequence of restructured models $M\rightarrow M_1 \rightarrow M_2....\rightarrow M_N$ to obtain a schedule of payments.

We say model has a \emph{cascade} structure, or simply model is a cascade, if set $J$ can be partitioned into sets $(J_1,J_2,...,J_m)$ such that the structure of obligations has the \flw \ form: there is a group of banks $J_1$ without any debts, there is a group of banks $J_2$ that owes money only to banks in $J_1$, etc, and, finally, there is a group of banks $J_m, m\leq n$, to which no bank owes money, and these banks may have debts to all other banks.

The structure of a cascade shows which banks are the main sources of debt, and where there are the weakest point in this chain of payments. The assumption that the initial model is sufficient is important. Otherwise, the process of mutual cancellations of debts inside of some subgroup of banks that lead to restructuring into a cascade, is impossible because of the strong assumption (A). Still, even in a deficient case the cascade can be used to find where the infusion of cash can alleviate the burden of obligations (see more in \citealp{Farthing2021}). We provide a fully worked-out example of a cascade restructuring in Subsection \ref{ssDDS}, using equations (\ref{pbc}) and (\ref{T}) that determine the discrete timings of status change. 

\begin{theorem}
\label{Th3} (a) Given \emph{any} model $(\mathbf{b},Q,\mathbf{c})$, there is at least one bank $j$ able to pay it debt without any money, i.e., having $p_j=b_j$, even with $c_j\leq 0$;

(b) Given any \emph{sufficient} model $(\mathbf{b},Q,\mathbf{c})$, it can be restructured into an equivalent cascade model $(\mathbf{b'},Q,\mathbf{c'})$.
\end{theorem}

The proof of Theorem \ref{Th3} is given by the Cascade Algorithm, based on the explicit construction presented in the next subsection.

\subsection{The Cascade Algorithm} \label{ssCAlg}

We provide a formal description of the algorithm in the proof of Theorem \ref{Th3}. Example \ref{Ex4} demonstrates the basic logic of the proof.

\paragraph{Proof of Theorem \ref{Th3}.}
If at time $0$ there is at least one paid-off bank, (a) is trivially true. Otherwise, there is an ergodic subclass $B$ with an \nvt \dst $\pi_i>0, i\in B$. Then by Lemma \ref{L53}, applying invariant rates $u_i=\pi_i$ to this subclass, we guarantee that all $c_i(t)$ remain equal to the initial values $c_i$. We also have that all $b_i(t)=b_i-tu_i$ are strictly decreasing. Let $t_i=\max_s\{s:b_i(s)>0\},$ $T_1=\min_{i}t_{i}$ and $i_1$ is the bank where this minimum is achieved. For simplicity, we assume that there is only one bank with this property; otherwise, bank $i_{1}$ will be replaced by a subset $J_{1}$. It means that $b_{i_1}(T_1)=0$, and this bank has no \lbls to all banks in $B$, and therefore to all banks in $J$.  All other banks in $B$ have debts $b_i(T_1)=b_i-T_1u_i=b_i^{1}>0$ to banks in $B$ only.

Let us denote the set of remaining banks $B\setminus i_1=S^1$ and represent $n-1$ dimensional vector of remaining debts of these banks, $\mathbf{b}^{(1)}=\mathbf{b}(T_1),$ as a sum of two vectors $\mathbf{b}^{(1)}=\mathbf{f}^{(1)}+\mathbf{e}^{(1)}$, where each component $\mathbf{f}^{(1)}_i$ is a remaining debt of bank $i$ to bank $i_1$, i.e., $f^{(1)}_{i}=b_i(T_1)q_{ii_1},$ $i\in S^1$. Then vector $\mathbf{e}^{(1)}$ is a $n-1$-dimensional vector of remaining \lbls of banks in $S^1$ with respect to each other. Let us obtain $n-1$ dimensional stochastic matrix $Q^{(1)}=\{q^{(1)}_{ij}\},$ $ i,j\in S^1$ by removing in $Q$ row and column $i_1$, i.e., all entries $q_{i_1i}$ and $q_{ii_1}$, $i\neq i_1$ and normalizing new rows, i.e., $q^{(1)}_{ij}=q_{ij}/\sum_{j\neq 1}q_{ij}$. If there is bank $i$ or a few banks in $S^1$ such that $q_{ii_1}=1$, they should be excluded from set $S^1$ and will be included later in set $J_2$. After such a modification of set $S^1$ all remaining banks form an ergodic subclass. Indeed, all banks left have a connection with some bank in this group. 

The next step is to obtain the invariant \dst $(\pi^1_i), i\in S^1$ for matrix $Q^{(1)}$. Then by Lemma \ref{L53}, applying invariant rates for this subclass, at some moment $T_2$ we obtain bank $i_2$ that paid its debts without changing its cash position. In other words, we apply the same step to set of banks $S^1$ with matrix $Q^{(1)}$ and debt vector $\mathbf{e}^{(1)}$ as we applied initially to set $J$ with matrix $Q$ and debt vector $\mathbf{b}$. If each moment $T_i$ only one banks pays its debt, then obviously in a $n-1$ steps we obtain a cascade with one bank at each level. Otherwise, in a smaller number of steps we obtain sets mentioned in Theorem \ref{Th3}.
$\qedsymbol$

\bigskip

\begin{example}\label{Ex4}
Let $n=4$,  matrix $Q$ be given by $q_{12}=q_{13}=q_{14}=\frac{1}{3},$ $q_{21}=\frac{1}{4},$ $ q_{23}=\frac{1}{2},$ $q_{24}=\frac{1}{4}$, $q_{31}=\frac{1}{6},$ $q_{32}=\frac{1}{2},$ $q_{34}=\frac{1}{3},$ $q_{41}=\frac{1}{6},$ $q_{42}=\frac{1}{2},$ and $q_{43}=\frac{1}{3}$.  The invariant \dst for the Markov chain defined by $Q$ is $\pi=\frac{1}{508}(84, 160, 147, 117)$. Consider the debt vector $\mathbf{b}=\left(33\frac{67}{78},56\frac{80}{117},55\frac{5}{39},20\right)$.
Using formula (\ref{cm}), we obtain $\mathbf{c}^{m}=\left(7\frac{1}{6},7\frac{5}{6},8\frac{5}{6}, -23\frac{5}{6}\right)$. 
\begin{eqnarray}
\label{MatrixFour}
Q=
	\begin{pmatrix}
         $0$ & \quad $1/3$ & \quad$1/3$ & \quad $1/3$ \\
        $1/4$ & \quad $0 $ & \quad$1/2$ & \quad $1/4$ \\
         $1/6$ & \quad $1/2$ & \quad$0$ & \quad $1/3$ \\
         $1/6$ & \quad $1/2$ & \quad$1/3$ & \quad $0$ \\
	\end{pmatrix}
\end{eqnarray}
Suppose that cash vector $\mathbf{c}\succ \mathbf{c}^m$, so our model is \emph{sufficient}. Our goal is to get to an equivalent cascade model. As in Example \ref{ExampleThree} and Lemma \ref{L53} on the first interval we are going use $u_i=\pi_i, i=1,...,4$. Then on this interval $c_i(t)=c_i$ for all $t$, and hence the moment $T_1=t_1$ is the first time when one of $b_i(t)$ hits zero. It is easy to check that this is bank $i=4$ and $t_1=\frac{20}{\pi_4}$. For $i=1,2,3,$ we have  $b_i(t_1)=20\frac{\pi_i}{\pi_4}:$ $b_1(t_1)=b_1-20\frac{84}{117}=19\frac{1}{2},$ $  b_2(t_1)=b_2-20\frac{160}{117}=29\frac{1}{3},$ $b_3(t_1)=b_3-20\frac{147}{117}=30,$ $ b_1(t_1)=b_1-20=0.$ 

\begin{figure}[t] 
\label{FigureB}
  \centering
  \includegraphics[scale=0.40]{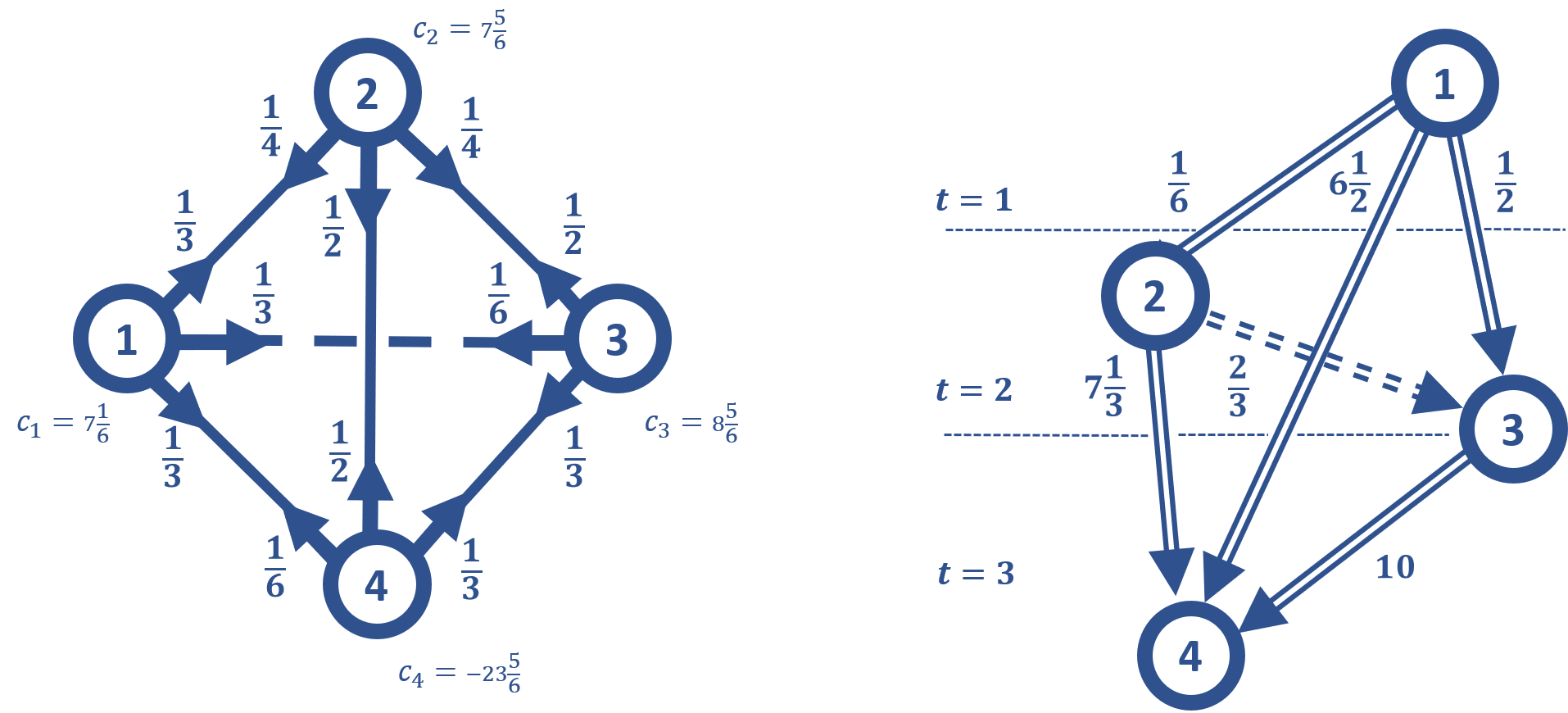} 
   \caption{(a) The initial matrix of relative liabilities with the minimum cash vector as the initial cash vector $\mathbf{c}$. (b) The cascade of payments in three steps.}
\end{figure}

Let us represent vector of remaining debts of all banks at time $t_1$ as the sum of two vectors, $\mathbf{b}(t_1)=\mathbf{f}^{(1)}+\mathbf{e}^{(1)}$, where vector $\mathbf{f}^{(1)}$ represents the vector of remaining debts of banks $1,2,3$ to bank $4$, and vector $\mathbf{e}^{(1)}$ represents the vector of remaining debts of banks $1,2,3$ to each other. We find these vectors using matrix $Q$. Thus, $f^{(1)}_{i}=b_{i}(t_1)q_{i4},$ and $e^{(1)}_{i}=b_{i}(t_1)-f^{(1)}_{i}$. We obtain $\mathbf{f}^{(1)}=\left(6\frac{1}{2}, \frac{22}{3}, 10\right)$ and $\mathbf{e}^{(1)}=(13, 22, 20)$. This is the debt vector from Example \ref{ExampleTwo}. So, we may consider sub-model with only three banks $1,2,3$ and with transition matrix $Q^{(1)}$ obtained from $Q$ by removing row 4 and column 4, and re-normalizing the new rows. 

As vector $\mathbf{e}^{(1)}=(13, 22, 20)$ was analyzed in Example \ref{ExampleTwo}, there is a cascade of the \flw \ form. At step 1, bank 1, having minimum cash $7\frac{1}{6}$, sends a check for $6\frac{1}{2}$ to bank 4, for $\frac{1}{6}$ to bank 2, and for $\frac{1}{2}$ to bank 3. At step 2, bank 2, having minimum cash $7\frac{5}{6}$, and $\frac{1}{6}$, send check for $\frac{2}{3}$ to bank 3, and $7\frac{1}{3}$ to bank 4. At step 3, bank 3, having minimum cash $8\frac{5}{6}$, with $\frac{1}{2}$ received form bank 1, and $\frac{2}{3}$ received from bank 2, sends check for $10$ to bank 4. Bank 4, having minimum cash $-23\frac{5}{6}$, received in total $6\frac{1}{2}+7\frac{2}{6}+10=23\frac{5}{6}$.  Figure 2(b) depicts the payments in this cascade. As it is always the case with a sufficient model, the payments can be uniquely identified by backward induction.

All banks paid their debts and their final cash positions are  $0$'s if $\mathbf{c}=\mathbf{c}^m$. If $\mathbf{c}\succ \mathbf{c}^m$, then it might be possible to settle all debts in less than three steps. 
\end{example}

\section{Extensions}
\label{SectionExtensions}
While our setup uses the initial setting of \cite{eisenberg2001systemic} as the
starting point, our model can be naturally generalized to work with many
extensions. For example, a modification of our algorithm can work with
liabilities or shares of different seniorities or liabilities with multiple maturities. In this Section, we briefly discuss possible extensions.

\subsection{Debt Seniority}
\label{Pex} %%%%%%%%%%%%%Pex
 First, we can assume that
matrix $Q$ is not obtained by the equalities $q_{ij}=b_{ij}/b_{i}$, and just
represents the priority of payments. Then of course we need the following
modifications. Now all out-pipes from tank $i$ are not closed simultaneously, but one by one, when the corresponding debt is paid, i.e., $(i,j)$ pipe is
closed when debt $b_{ij}$ is paid. Corresponding equations and the times of
status change can be easily modified. If we assume that some debts should be
paid not just faster but before other payments, then at the initial moment
not all $(i,j)$ pipes are open but only for $j$ in the senior (for $i$)
class. When these debts are paid, then the other group of $(i,j)$ pipes is
open, etc.

Similarly, while we assumed, that all \textquotedblleft
positive\textquotedblright\ banks have the same total out-capacities equal
to one, the analysis is easily extended to heterogeneous rates (if the
regulator considers it important to prioritize some payments). Note that,
although the dynamics of the continuous-time model will be changed, the
clearing vector will be the same if proportionality of all payments will be
the same. Of course, if the requirement of proportionality of payments is
changed, then not only will the dynamics of the continuous-time model be
changed, but the clearing vector as well.

The next observation is that because of the irreversibility property not all
banks can leave group $J_{+}(0)$ before $T_{\ast }$, we can have a raw
estimate $T_{\ast }\leq \max_{i}b_{i}$. If the total capacity is changed
from $1$, as in our basic setup, to $m$ then $T_{\ast }$ will be changed to $%
T_{\ast }/m$.

Let us discuss the following potential question: \emph{how much extra money $%
x_{i}\geq 0$ should be given to each (or some) bank to avoid all defaults?}
Note that the sum of unpaid debts, i.e., $\sum_{i\in J}k_{i},\ \
k_{i}=b_{i}-p_{i}(T_{\ast })$ can be substantially more than $X=\sum_{i\in
J}x_{i}$. Let us consider the following mechanism. Change the initial cash
positions of all banks in $J_{0}(T_{\ast })$ from $c_{i}$ to $c_{i}^{\prime
}=c_{i}+k_{i}$, and run the continuous-time model again. Intuitively it is
clear and can be proved that with these new initial positions, all debts
will be paid and it is possible that some of these bank will have at the end
positive cash positions $c_{i}(T_{\ast }^{\prime })$. Intuitively it is
clear and can be proved that if now the initial cash positions of all banks
in $J_{0}(T_{\ast })$ are changed from $c_{i}$ to $c_{i}^{\ast }=c_{i}+x_{i}$%
, where $x_{i}=k_{i}-c_{i}(T_{\ast }^{\prime })$, e.g., by a central bank or
a private consortium, then all these banks will finish with zero cash
positions but, at the same time, fully paying their debts. Obviously $0\leq
x_{i}\leq k_{i}$. This problem is related also to the distinction between
``real defaults" vs ``temporarily defaults".
%%%%%%%%%%%%%%%%%%%%%  `` left "

\subsection{Multiple Maturities}
\label{VA}
In a recent paper, \cite{kusnetsov2019interbank} analyze the case of multiple maturities. Our model allows a straightforward extension to incorporate this possibility. Suppose that there is a scale of time $[0,+\infty)$ and each bank may have obligations with possibly different due dates, i.e., all liabilities have extra maturity time (stamp) $b_ij(t_k), k=1,2,..., n_i$. As in \cite{kusnetsov2019interbank}, we start with two due dates for all banks, short and long, $t_1<t_2$. We modify our model as follows.  All parameters of a system of banks are ``doubled", have an extra index 1 or 2, like $b_i(s)=\sum_jb_{ij}(s), s=1,2$. All banks have two types of tanks, \emph{basic} (1) and \emph{dormant} (2), two cash positions $c_i(s), s=1,2,$
and two types of out-going and in-going pipes, also marked as short (1) and
long (2). The dormant tank of bank $i$ represents an account that can be
used only starting from long moment, and which accumulates money from other
banks that have short liabilities to $i$. All ``short" pipes are directed to
short tanks, all ``long" pipes are directed to long tanks. 

To modify capacities, we will have to introduce two new axioms. The first, similar
to \cite{kusnetsov2019interbank}, stems from the requirements of the UK insolvency rules is: All liabilities with different maturities are treated with the same priority within the same seniority class. It might be reasonable to introduce a second assumption, different from \cite{kusnetsov2019interbank}, where ``a bank that is liquidated under the insolvency rules ceases to exist and cannot recover even if liquidators recover sufficient assets to fully compensate all creditors." Specifically, one can require  that all banks that are ``short bankrupt'', \emph{must} borrow money with fixed interest from outside banks to pay for their short obligations if their long obligations from other banks will allow them to finish avoiding bankruptcy in the long run. This would allow to extend our analysis to the environment with multiple maturities. 

\subsection{Clearing Equation as the Bellman Optimality Equation}

In our continuous-time model, we used the mathematical technique of Markov chains to analyze the multiplicity of solutions in a deterministic dynamical system. In fact, there is a deeper relationship between the Eisenberg and Noe problem and some classic problems in the theory of Markov chains. The clearing equation (\ref{Basic}) has strong resemblance to the nonlinear equation 
\begin{equation}
    \mathbf{v}=\max (\mathbf{c}+Q\mathbf{v},\mathbf{b}), \label{EquationBellman}
\end{equation}
 used in many applications that use Bellman optimization. Most linear models, e.g., the Leontief closed and open models with extra constraints will satisfy (\ref{EquationBellman}). 
 
 Consider the classic problem of \emph{optimal stopping of a Markov chain} with transition matrix $Q$, where a decision maker observing the chain has, at each moment in time,  two options, either to continue or to stop (\citealp{puterman2014markov}). In such a setting, the Bellman (optimality) principle takes the form of equation (\ref{EquationBellman}), where vector $v=(v_{i})$ is the value function, i.e., $v_{i}$ is the maximal possible expected reward over all possible stopping times if the Markov chain starts at state $i$, $c=(c_{1},...,c_{n})$ is a vector that consists of \textit{current rewards }$c_{i}$ in state $i$, and $b=(b_{1},...,b_{n})$ is a vector of \textit{terminal rewards,} where $b_{i}$ is the terminal reward if the Markov chain stops at state $i$. Note that in the Bellman equation maximum is taken instead of minimum and the straight matrix $Q$ is used instead of the transposed one. 
 
 A simple recursive algorithm, the \emph{State Elimination Algorithm}, to solve the Bellman equation (\ref{EquationBellman}) was developed in \cite{sonin1999state,sonin2006optimal} and was modified to calculate the classic and generalized Gittins indices in \cite{sonin2008generalized}. The relationship between the Eisenberg and Noe basic equation (\ref{Basic}) and the optimal stopping problem defined by (\ref{EquationBellman}) is discussed in \cite{kabanov2018clearing}.

\section{Conclusion\label{CON}}

In this paper, we develop a continuous-time model of clearing in financial
networks. This approach provides an intuitive and simple recursive solution
to a classical static model of financial clearing introduced by \cite{eisenberg2001systemic}. The same approach provides a useful tool to solve nonlinear equations involving a linear system and max min operations similar to the
Bellman equation for the optimal stopping of Markov chains and other
optimization problems. Allowing financially constrained banks to repay their debts at the maximum-available speed without any liquidity injections or guarantees will result in a clearing payment vector. Our results show that, at
least theoretically, there is no need in detailed regulation in a situation
of financial distress if the mechanism of resolving simultaneous payments is
set right. On the other hand, our model provides a convenient tool to study the optimal strategy of a central agency to minimize the potential contagion effect
triggered by failure of some banks. Finally, our approach allows to resolve simultaneous
or nearly simultaneous payments in a time of financial distress by allowing
to \textquotedblleft stretch" an instant moment into a finite time interval.

\pagebreak
\bibliography{TanksReferences}

\pagebreak

\section*{Appendix}
\addcontentsline{toc}{section}{Appendix}

\renewcommand{\theequation}{\mbox{A\arabic{equation}}}

\renewcommand{\thesection}{\mbox{A\arabic{section}}}

\renewcommand{\thesubsection}{\mbox{A\arabic{subsection}}}

\renewcommand{\thetheorem}{\mbox{A\arabic{theorem}}}

\renewcommand{\theexample}{\mbox{A\arabic{example}}}

\setcounter{equation}{0}

\setcounter{section}{1}

\setcounter{subsection}{0}

\setcounter{theorem}{0}

%\addtocontents{toc}{\protect\setcounter{tocdepth}{1}}
\subsection{Proof of Theorem \ref{Main} (b)}
%\addtocontents{toc}{\protect\setcounter{tocdepth}{2}}
\label{ss81}

In this subsection we prove the monotonicity property of rates, which is equivalent to point (b) of Theorem \ref{Main}.
We prove by induction on $k$, where $k$ is the number of the interval $\Delta _{k}=[T_{k},T_{k+1}),k=1,2,...$. We know that generally group $J_{0}(t)$ is changing in time, can be increased, decreased, become empty set and appear again. When $k=1$,
$u_{i}=1$ for all $i$ and as we explained in Subsection \ref{ssEDS}, at the moment of the first appearance of
the group $J_{0}=\{j\}$ the value of out-rate for the member $j$ of this group, $v_j=u_j<1$. Suppose this statement
is true on the status interval $\Delta_{k},k>0$. As before we skip further indication to $k$, denoting $u_{i,k}=u_{i},\Delta
_{k}=\Delta $, $T_{k}=T$, $T_{k+1}=T^{\prime }$.
%, and partitions $P_{k}=P,P_{k+1}=P^{\prime }$. ,Q^T_{0}=Q^T_{J_{0,k}}
We denote other vectors, coordinates, elements of partition and matrices before status change, i.e., on the interval \ $\Delta_{k}$, as $J_0=J_{0,k}$ and the
%%% $b^{\ast },c^{\ast },P^{\ast },Q^{\ast },Q_{0}^{\ast }$ ,Q'_{0}=Q^T_{J_{0,k+1}}
values after change as $J'_0=J_{0,k+1}$, etc. We denote the solution of the
equilibrium equation on the interval $\Delta _{k}$ as $(v_{i})$, and on the interval $\Delta _{k+1}$ as $(w_{i})$.
 Thus, our induction statement is $0<v_{i}<1, i\in J_0$, and our goal is to prove that $v_{i}\geq w_{i}$ for
all $i\in J'_{0}$.

The status change at moment $T_{k}=T$ occurs when $b_{r}(T)=0$ or $c_{r}(T)=0$, where $r\in J_+\cup J_0$.
If $b_{r}(T)=0$ for $r\in J_+$, this means that zero group is unchanged but the previous input vector $(e_{i})$ can be decreased if bank $r$ was a sponsor of a zero group. Then according to point (c) of Lemma \ref{Lemma51} the new solution of an \eqm equation can be only strictly decreased. More difficult are cases when $c_r(T)=0$ for $r\in J_{+}$ and when $b_r(T)=0$ for
$r\in J_0$. In the first case $J'_0=J_0\cup r$, and if $|J_0|=m$, the new system has $m+1$ variables, with value $u_r=1$ changed to an (extra unknown) value $w_{r}$. In the second case, the new \eqm system has only $m-1$ variables.
%%%%%%%%%%%%%%% large !! \large {
We prove only the first case, the second case can be treated in a similar way. We consider even (slightly) more general statement for \flw \ situation that includes our first case: suppose we have two partitions:
an initial $P=(J_{+},J_{0},J_{\ast })$, such that positive banks have a subset $L$ with $n_i<1, i\in L$ and
new \prt $P'=(J'_+=J_{+}\setminus L,J'_{0}=J_{0}\cup L,J_{\ast })$.

\begin{lemma}
\label{LA1} Let under assumptions above $v_i=n_i$ be the solution of \eqst for \prt $P$, and $w_i=n'_i$ be the solution of \eqst for \prt $P'$. Then $v_i\geq w_i$ for all $i\in J'_{0}$, and $n_i\geq n'_i$ for all $i$.
\end{lemma}

A heuristic proof is as follows. When banks in $L$ are classified as positive, their out-rates are equal to one, i.e. maximum possible; when they are included as  members of zero group, the input vector into zero group is diminished and since all \blgs \ are the same, as a result, the in-rates (out-rates) in a new zero group become lower. Then all new in-rates are lower. The technical difficulty in a rigorous proof is that we have to compare the solutions of two \eqst of different size. It is sufficient to consider only the case when $|L|=1$. Let $L=\{r\},$ $n_r<1$.
%%%%%%%%%%%%%%%%%%

\begin{proof}
The equilibrium system (\ref{int}) for $w_{i}$ with $J'_{0}=J_{0}\cup r$, is
\begin{eqnarray}
w_{i}&=&e'_{i}+\sum_{j\in J_{0}}w_{j}q_{ji}+w_{r}q_{ri},\quad i\in J_{0}, \notag \\ 
w_{r}&=&e'_{r}+\sum_{j\in J_{0}}w_{j}q_{jr}, \label{wi}
\end{eqnarray}
where $e'_{i}=\sum_{j\in ({J_{+}\setminus r)}}q_{ji}=\sum_{j\in {J_{+}}}q_{ji}-q_{ri}=e_i-q_{ri}$ and $e'_{r}=\sum_{j\in {J_{+}}}q_{jr}=e_r$.

The system for $v_{i},$ $i\in J_0$ is: $v_{i}=e_{i}+\sum_{j\in J_{0}}v_{j}q_{ji}.$
%\vspace{.2cm}
We can rewrite this system to make it similar to (\ref{wi}), using artificial variable $v_{r}=1$, and equalities $e'_{i}=e_i-q_{ri}$ and $e_{r}=e'_{r}$ as:
\begin{equation}
v_{i}=e'_{i}+\sum_{j\in J_{0}}v_{j}q_{ji}+v_{r}q_{ri},i\in J_{0};\ \ \
1=v_{r}=e'_{r}+a_{r}+\sum_{j\in J_{0}}v_{j}q_{jr},  \label{vi}
\end{equation}
%% \sum_{j\in J_{+}}q_{jr}      \sum_{j\in J_{0}}v_{j}q_{jr}, $c_{r}(T)=0$ from $c_{r}(t)>0$ for $t<T$.
where $a_{r}=1-(e'_{r}+\sum_{j\in J_{0}}v_{j}q_{jr})$. We know that this system has a solution $v_i<1, i\in J_0, v_r=1$. Now we want to compare solutions of systems (\ref{wi}) and (\ref{vi}) to show that $v_i\geq w_i, i\in (J_0\cup r)$ and $w_{r}<1$. To use point (c) of Lemma \ref{Lemma51}, we need to show only that $a_{r}\equiv 1-(\sum_{j\in J_{+}}q_{jr}+\sum_{j\in J_{0}}v_{j}q_{jr})>0$. We assumed that $n_r=n_{r}(T_-)<1$. But
%%%%%%%%%%%  Xr  Now we can recall that the status of $r$ was changed from $+$ to $0$ not without a reason. It was changed because the cash position of bank $r$ has reached zero from above. Hence, by formula (\ref{pbc}), using $u_{r}(T_-)=1$, we have balance rate $d_{r}(T_-)=n_{r}(T_-)-1<0$. Thus, we have
$n_{r}(T_-)=\sum_{j\in J_{+}}q_{jr}+\sum_{j\in J_{0}}v_{j}q_{jr}=1-a_r<1$ and therefore $a_{r}>0$. Then, using point (c) of Lemma \ref{Lemma51}, we obtain that $v_{i}\geq w_{i}$ for all $i\in J_{0}$ and that $1> w_{r}$, i.e. $w_{i}<1$ for all $i\in J'_{0}$.
Since out-rates in positive group remain the same and out-rates in new zero group, including new member $r$, are lower, then all in-rates are lower. \par
%%%%%%% -q_{ri}=e^*_{i}-q_{ri}, using Lemma \ref{Lemma1}, we obtain $y_{i}\geq 0$ for all $i\in (J_{0}\cup r)$, i.e., part (c) of Theorem \ref{Main} holds in case 2).
The second case, when $b_r(T)=0, $ $r\in J_0$, and the new \eqm system has only smaller variables can proved similarly. Thus, we proved part (b) of Theorem \ref{Main}, or, equivalently, the monotonicity property for in-rates and out-rates.
\end{proof}

\subsection{The Big Bang Effect}\label{SubsectionBigBang}
The initial moment of time, $t=0$, is exceptional if at this moment set $J_{0}(0)$ is not empty. The reason for that is as follows. If at time $t=0$ all banks have a positive amount of cash, then $u_{i}(t)=1$ for all banks on some interval $[0,T_{1})$, where moment $T_{1}$ is the first moment when one of the banks either pays its debt or its cash position hits zero level. After that, we can apply the monotonicity property, i.e., part (b) of Theorem \ref{Main}. However, if at time zero there are some zero banks then, informally, there is a Catch-22 situation. Some zero banks at moment $t=0$ should be classified as positive if the input rates for them exceed one. This means that at the \textquotedblleft next moment after zero" they instantly become positive. Equivalently, for these banks the balance rate $d_{i}(0)>0$, or equivalently $c_{i}^{\prime }(t)_{t=0}>0$. But to find the input rates for all banks by solving the equation (\ref{int}) we need to know which banks are positive and which are in the zero group. In some cases, like in our Example \ref{ExampleOne}, where only one bank had the initial cash equal zero, and input-rate more than one, it was easy to reclassify this bank as positive. If there are two or more zero banks than such reclassification cannot be done so easily. One way to circumvent this problem is as follows.

Assume that all zero banks received a small positive amount $\varepsilon $ prior to zero moment and then, during the small \textquotedblleft probe"\ time interval $\Delta _{0}=[0,t_{m})$,
having length of order $\varepsilon $, we have a regular case analyzed in Subsection \ref{SubsectionRegularCase}. Then during this interval each of the initially zero
banks will \textquotedblleft reveal" its real status: the cash values of
\textquotedblleft real-zero"\ banks, having balance rates $d_{i}(t)$
negative, very soon hit zero level during a series of close moments $%
t_{1},...,t_{m}$, with $m$ potentially equal zero. \textquotedblleft
Real-positive" banks will remain positive on the small interval $(0,t_{m})$,
and will remain positive at least until a more distant moment $T_{1}$. This
means that on a very short interval, we may have moments $%
t_{1},t_{2},...,t_{m}$ of fast status changes, and at $t_{m}$ all the real
statuses are revealed.
\end{document}